\documentclass[onecolumn,aps,longbibliography,superscriptaddress]{revtex4}
\usepackage{graphicx}  
\usepackage{graphics}
\usepackage{dcolumn}   
\usepackage{bm}        
\usepackage{amssymb}   
\usepackage{color}
\usepackage{natbib}
\usepackage{epstopdf}
\usepackage{epsfig}

\begin{document}
\title{Stochastic modeling of phenotypic switching and chemoresistance in cancer cell populations}
\author{Niraj Kumar}
\affiliation{Department of Physics, University of Massachusetts Boston, Boston MA 02125, USA}
\author{Gwendolyn M. Cramer}
\affiliation{Department of Physics, University of Massachusetts Boston, Boston MA 02125, USA}
\affiliation{Current address:Department of Radiation Oncology, Perelman School of Medicine, University of Pennsylvania, Philadelphia, PA}
\author{Seyed Alireza Zamani Dahaj}
\affiliation{Department of Physics, University of Massachusetts Boston, Boston MA 02125, USA}
\affiliation{Current address: School of Physics, Georgia Institute of Technology, Atlanta GA 30332, USA}
\author{Bala Sundaram}
\affiliation{Department of Physics, University of Massachusetts Boston, Boston MA 02125, USA}
\author{Jonathan P. Celli}
\affiliation{Department of Physics, University of Massachusetts Boston, Boston MA 02125, USA}
\author{Rahul V. Kulkarni}
\affiliation{Department of Physics, University of Massachusetts Boston, Boston MA 02125, USA}
\begin{abstract}
Phenotypic heterogeneity in cancer cells is widely observed and is often linked to drug
resistance.  In several cases, such heterogeneity in drug sensitivity of 
tumors is driven by stochastic and reversible acquisition of a drug
tolerant phenotype by individual cells even in an isogenic
population. Accumulating evidence further suggests that cell-fate transitions such as the epithelial
to mesenchymal transition (EMT) are associated with
drug resistance. In this study, 
we analyze stochastic models of phenotypic switching to provide a framework for
analyzing cell-fate transitions such as EMT as a source of
phenotypic variability in drug sensitivity. 
Motivated by our cell-culture based experimental
observations connecting phenotypic switching in EMT and drug
resistance, we analyze a coarse-grained model of phenotypic switching
between two states in the presence of cytotoxic stress from
chemotherapy.  We derive analytical results for time-dependent
probability distributions that provide insights into the
rates of phenotypic switching  and characterize initial phenotypic 
heterogeneity of cancer cells. The results obtained can also shed light on fundamental questions
relating to adaptation and selection scenarios in tumor response to
cytotoxic therapy.\\
\end{abstract}
\maketitle
\section{Introduction}
\noindent Acquisition of drug resistance constitutes a major challenge in cancer therapy
\cite{housman2014drug,gottesman2002mechanisms, chisholm2016cell, bozic2017resisting, 
pogrebniak2018harnessing, nikolaou2018challenge, salgia2018genetic, dagogo2018tumour,
zhou2009tumour, zahreddine2013mechanisms, 
holohan2013cancer, garraway2012circumventing,shaffer2017rare}. Therapeutic agents (with widely varying
biochemical mechanisms) often exhibit a common pattern of providing an
initial reduction in tumor burden followed by recurrence of
therapeutically resistant disease with more aggressive
progression  \cite{nikolaou2018challenge, 
holohan2013cancer, garraway2012circumventing, chabner2005chemotherapy}.
Tumor recurrence, which is a major obstacle for cancer cure, is
primarily associated with the survival and growth of cell phenotypes that are
resistant to chemotherapy \cite{gatenby2018evolution,gallaher2018spatial,castorina2009tumor}. Therefore, in order to develop new
strategies for the effective treatment of human cancers, a
quantitative understanding of the underlying processes leading to
drug resistance is essential.\\

\noindent Cellular phenotypic heterogeneity is widely observed in many cancers \cite{pardal2003applying, chisholm2016cell, pogrebniak2018harnessing,dagogo2018tumour,meacham2013tumour,marusyk2012intra} as a tumor is often composed of multiple subpopulations \cite{gupta2009cancer,zhou2014nonequilibrium} that show different responses to chemotherapy \cite{zhou2009tumour}.
In particular, cellular phenotypes that are not sensitive to drugs survive the treatment and can drive drug resistance.
As the underlying processes that can lead to the emergence of resistant cells are often stochastic, tumors may locally  contain varying 
numbers of resistant cells. Therefore, quantifying the statistics of drug resistant cells in a tumor is important for effective therapy.
Specifically, we are interested in studying population heterogeneity at the start of therapy and aim to 
address an important issue of therapeutic importance, namely, how to quantify randomness in the numbers of resistant cells prior to drug treatment.\\

\noindent In analyzing population heterogeneity in tumors, a basic question that arises is: How are cell phenotypes that confer survival advantage in the 
presence of chemotherapeutic drugs generated? A common explanation for the emergence of such phenotypes revolves around
Darwinian selection of pre-existing cellular heterogeneity that arises due to random genetic mutations \cite{nowell1976clonal,sottoriva2013intratumor, burrell2013causes}.
However, the fact that resistant cells switch reversibly to sensitive cells, and that resistant cells often appear on short time intervals (hours to few days), starting from clonal populations, suggests that non-genetic factors play a major role
in the generation of phenotypic heterogeneity \cite{pisco2013non, pisco2015non,pogrebniak2018harnessing,salgia2018genetic,brown2014poised,shaffer2017rare, su2017single,inde2018impact}. Such non-genetic phenotypic heterogeneity can arise due to multistability in the underlying gene expression dynamics \cite{chang2008transcriptome, huang2005cell} and noise in gene expression\cite{kaern2005stochasticity}. \\

\noindent These observations suggest that there are two distinct, though not mutually exclusive, mechanisms for the onset of  
drug resistance in cancer cells: 1) cell phenotypes that are resistant to chemotherapy pre-exist in the 
tumor prior to treatment and are selected for during the treatment, and 2) cells are induced to develop
or acquire resistance due to treatment. In the literature, the former scenario is termed selection  while later corresponds to adaptation.
This adaptation-selection scenario was first explored in the famous Luria-Delbr{\"u}ck experiments \cite{luria1943mutations} 
to understand the mechanism of bacterial resistance to bacteriophage infections. The corresponding analysis gave rise to the celebrated fluctuation test which is also used to estimate  mutation rates in bacteria. It is important to note that, while in the Luria-Delbr{\"u}ck case phenotypic changes are driven by genetic mutations and thus an irreversible process, in our study,  we are considering phenotypic changes that are reversible. Besides reversible phenotypic switching, it is important to consider intrinsic stochasticity in the underlying processes and to characterize cellular heterogeneity as highlighted by  previous studies focusing on modeling drug resistance in cancer \cite{kessler2014resistance,komarova2006stochastic}. \\

\noindent In consideration of cellular mechanisms likely to be associated with drug resistance, the epithelial-mesenchymal transition (EMT) emerges as a logical candidate. EMT is a conserved cellular program that enables cells of epithelial lineage to transiently  acquire traits of mesenchymal cells, including reversible loss of adherens junctions and gain of proteins associated with enhanced motility, adhesion to extracellular substrates and remodeling of the extracellular matrix \cite{kalluri2009basics,lamouille2014molecular}. In cancer cells, this ability to reversibly adopt a more motile phenotype has been linked to tumor invasion and metastasis \cite{heerboth2015emt,yang2008epithelial, zhang2018epithelial} but, more importantly for this study, EMT is also directly linked to chemotherapy resistance and cancer stem cell (CSC) properties \cite{ thiery2009epithelial,singh2010emt}. In the context of this background, experimental studies described herein focus on established markers of epithelial and mesenchymal phenotype in relation to chemotherapy response, which in this report involves pancreatic ductal adenocarcinoma (PDAC) cells. While recognizing that EMT is more likely a spectrum of intermediate states \cite{lu2013microrna,jolly2015implications, jolly2016stability,hong2015ovol2,li2018landscape}, the strong correlation in phenotype and drug response reported here motivates the adoption of two coarse-grained states to be used in the model development. Specifically,  a relatively drug-sensitive state with more pronounced epithelial characteristics (E); and a drug-resistant state with increased mesenchymal characteristics (M). In the following sections, we will consider these phenotypes to form the basis of a two-state model of the dynamics of phenotypic switching and associated survival of cancer cells under cytotoxic stress\cite{pisco2013non,zhou2014nonequilibrium}.\\


\noindent The paper is organized as follows. In Section II, a set of motivating experiments is described, in which drug resistance is evaluated as a determinant of phenotype in pancreatic cancer cells in vitro, and conversely, phenotype as a determinant of drug response in the same cells. Motivated by these studies, a two phenotype switching model is described in Sec III. An analytic approach 
to quantify population heterogeneity at the start of therapy is presented in the Sec IV. Then, a 
protocol for estimating switching parameters is presented in 
Sec V. In Sec VI, we present an approximate approach for characterizing the probability distribution of the 
fraction of resistant cells in a population followed by conclusions presented in Sec VII. 

\section{Evaluation of drug resistance and phenotype in cell culture studies}

\noindent We first sought to compare phenotypic traits in naive and drug-resistant pancreatic ductal adenocarcinoma (PDAC) cells. PANC1 cells (a quasimesnchymal PDAC cell line \cite{collisson2011subtypes}) were exposed to increasing doses of oxaliplatin chemotherapy over successive passages until resistant cells were stable through multiple passages and cryopreservation.
As shown in Figure 1 (upper panels), acquisition of chemoresistance leads to a marked change in phenotype from naive cells displaying characteristically epithelial adherens junctions, to drug-resistant cells with highly branched morphology, no evident E-cadherin, and marked increase in cytoskeletal vimentin (IF quantification, upper right). This pattern of changes in E-cadherin and vimentin expression are classic and well-established markers of EMT.  We further examined the reciprocal scenario, in which the same parental cells were directly induced \cite{kalluri2009basics}to undergo EMT via administration of exogenous TGF-$\beta$ (Figure 1, lower panels).
The resultant phenotype is strikingly similar to that of our drug-resistant cells and importantly, exhibits resistance to chemotherapy similar to when resistance was acquired directly through drug exposure. Collectively these results display a symmetry in that acquisition of drug resistance in epithelial cancer cells leads to increase in mesenchymal characteristics, while direct transition from epithelial to mesenchymal phenotype leads to drug resistance. \\
\begin{figure}[h]
	\includegraphics[bb=23 131 757 459,width=15cm]{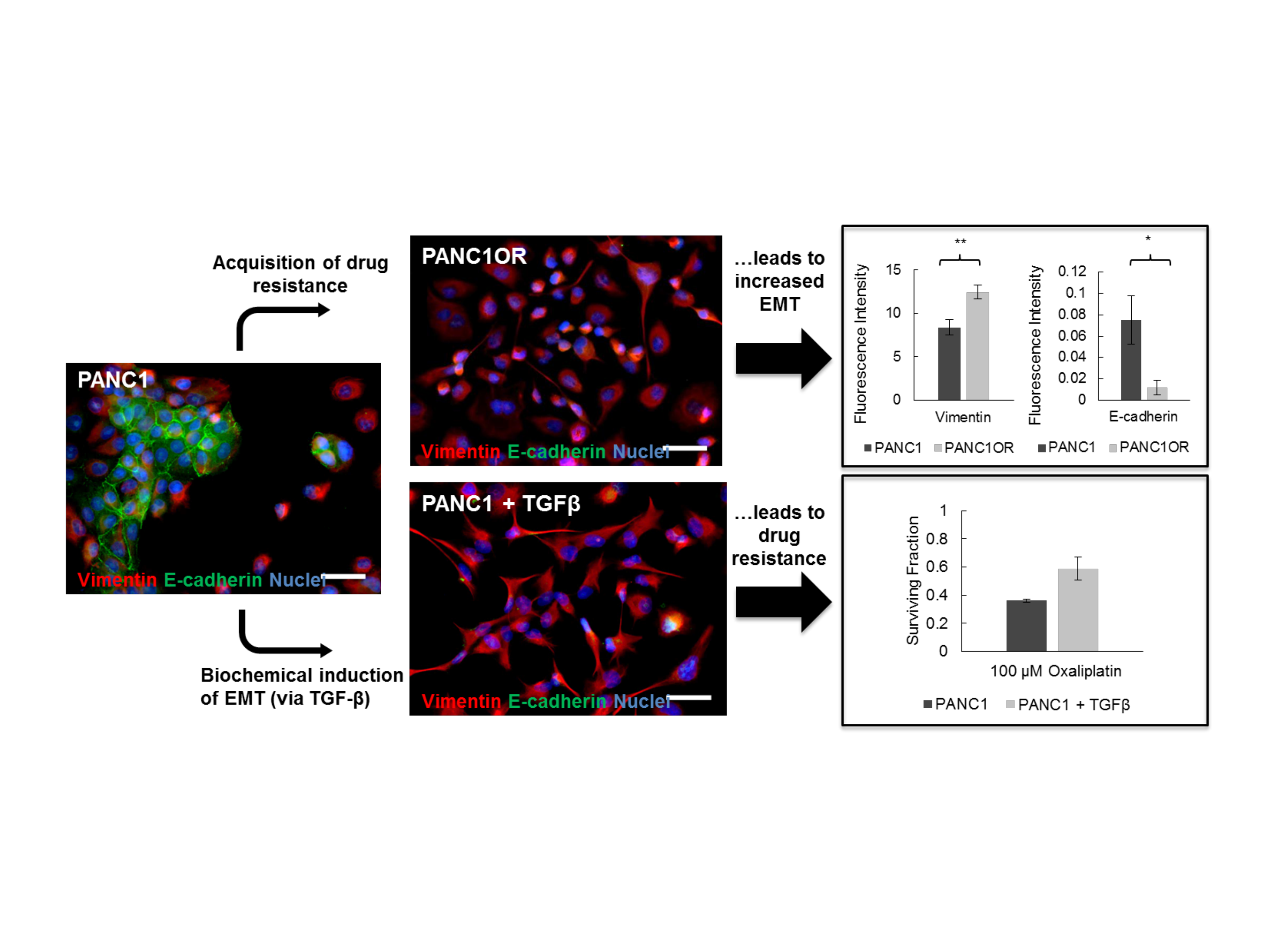}
	\caption{Equivalence in the acquisition of chemotherapy resistance and epithelial-mesenchymal transition in pancreatic cancer cells.}
\label{fig:expt}
\end{figure}

\subsection*{Methods}
\noindent{\bf Cell culture and reagents:} PANC1 cells were obtained from the American Type Culture Collection (Manassas, VA), and grown in T-75 cell culture flasks according to ATCC guidelines. DMEM medium (HyClone; Logan, UT) was supplemented with 10\% FBS (HyClone; Logan, UT), 100 IU/mL penicillin and 1\% streptomycin (HyClone; Logan, UT), and 0.5 ug/mL Amphotericin B (Corning; Corning, NY). The drug-resistant subline, PANC1-OR was generated as described previously \cite{cramer2017ecm}. Briefly, increasing concentrations of oxaliplatin were added to each cell type in regular media over the course of approximately 25 passages until a stable proliferative phenotype without chemotherapy was observed and maintained following cryopreservation and confirmed by comparative dose response and measurement of a statistically significant increase in IC50.\\

\noindent{\bf Immunofluorescence sample preparation and imaging:} Formaldehyde-fixed cells in optical-bottom multiwell plates were incubated overnight at $4^{\circ}$C with primary antibodies against e-cadherin and vimentin (Cell Signaling EMT Duplex; Danvers, MA). After washing with PBS, cells were incubated for 1 hour with mouse or rabbit Alexa Fluor secondary antibodies (Cell Signaling; Danvers, MA). Cells were mounted with ProLong Gold Antifade reagent containing DAPI (ThermoFisher Scientific Molecular Probes; Waltham, MA) and imaged after 24 hours using a Zeiss LSM 880 confocal microscope with the same detector settings and excitation laser power settings across groups. Images were analyzed using custom Matlab scripts where fluorescent signal for each protein was normalized to the number of cells based on DAPI-stained nuclei. \\

\noindent{\bf Therapeutic response assessment:} In sample wells receiving chemotherapy treatment, oxaliplatin (Selleck Chemical; Houston, TX) was added to the media at doses ranging from 0.1 to 500 $\mu$M for 48 hours. In experiments where EMT was induced via TGF-beta, 10 ng/mL human recombinant TGF-beta (Gibco, Thermo Fisher Scientific) in 1\% FBS DMEM was added to designated wells for 48 hours and respective comparison groups were also grown in 1\%  FBS for the same duration. In all therapeutic studies treatment conditions were prepared in at least triplicate within each batch including internal controls with sham manipulations. Therapeutic response was assessed via the CellTiter 96® AQueous One Solution Cell Proliferation Assay (Promega; Madison, WI) at 490nm absorbance in a BioTek® Epoch Microplate Spectrophotometer.\vspace{2cm}\\

\section{Coarse-grained Model}

\noindent  Motivated by the preceding observations and by previous work \cite{pisco2013non,zhou2014nonequilibrium}, we now consider a simple coarse-grained model (Fig. 2) for phenotypic heterogeneity in tumor cells. We consider that the population of cancer cells consists of two distinct subpopulations; drug-sensitive or drug-tolerant.  Based on our experimental results, we denote the
drug-sensitive population by $E$ (for epithelial phenotype) and the drug-tolerant population by $M$ (for mesenchymal phenotype).  
The processes that control the evolution of tumor heterogeneity are as follows: 1) \emph{birth}: each $E$-type 
or $M$-type cell gives rise to birth of new cells of the same type with rates $k_{E}$ and $k_{M}$,
respectively; 2) \emph{death}: each $E$-type ($M$-type) cell degrades with rates $\mu_E$  ($\mu_M$); 3) \emph{phenotypic switching}: an $E$ cell can switch to a $M$ cell with rate $k_{EM}$, and $M$  cell can switch back to $E$ cell with rate $k_{ME}.$\\

\noindent At any time $t$, the state of the system is defined by the number of $E$ and $M$ cells. The temporal evolution of the corresponding probability distribution is given by the master equation:
\begin{eqnarray}{\label{eMaster}}
 \frac{\partial P(E,M,t)}{\partial t}&=&k_E(E-1) P(E-1,M,t)+k_M(M-1)P(E,M-1,t)\nonumber\\&+&\mu_E(E+1)P(E+1,M,t)+\mu_M(M+1)P(E,M+1,t)\nonumber\\&+&k_{ME}(M+1)P(E-1,M+1,t)
                                       +k_{EM}(E+1)P(E+1,M-1,t)\nonumber\\&-& \left[k_E E+k_MM+\mu_E E+\mu_M M+k_{ME} M+k_{EM}E\right]P(E,M,t),
\end{eqnarray}
where $P(E,M,t)$ denotes the probability that there are $E$ and $M$ numbers of epithelial and messenchymal cells present at time $t$. \\

\begin{figure}[h]
\centering
\includegraphics[width=9cm]{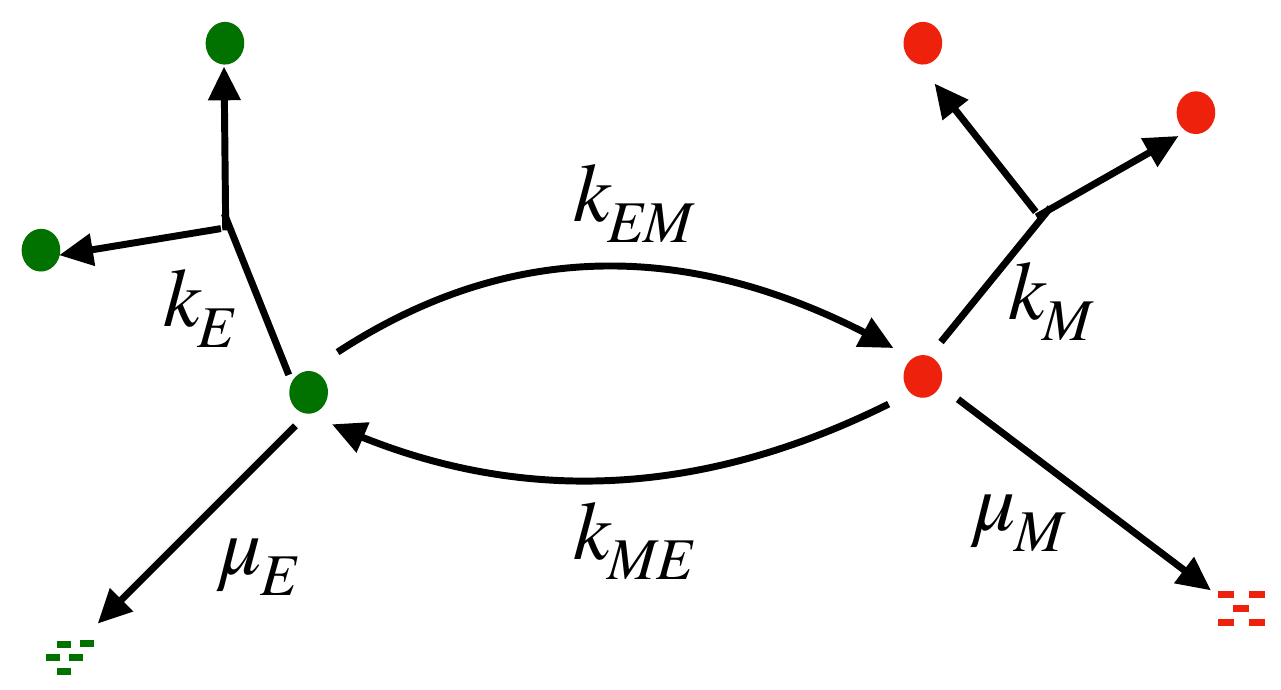}	
\caption{Schematic representation of the two phenotype EMT model of tumor growth:  Sensitive and resistant phenotypes are shown as green and red circles respectively. The phenotypic switching rates are represented by 
$K_{EM}$ and $K_{ME}$, birth rates by $k_E$ and $k_M$, and death rates by $\mu_E$ and $\mu_M$. }
\label{fig:model}
\end{figure}

\begin{figure}[h]
\centering
\includegraphics[width=9cm]{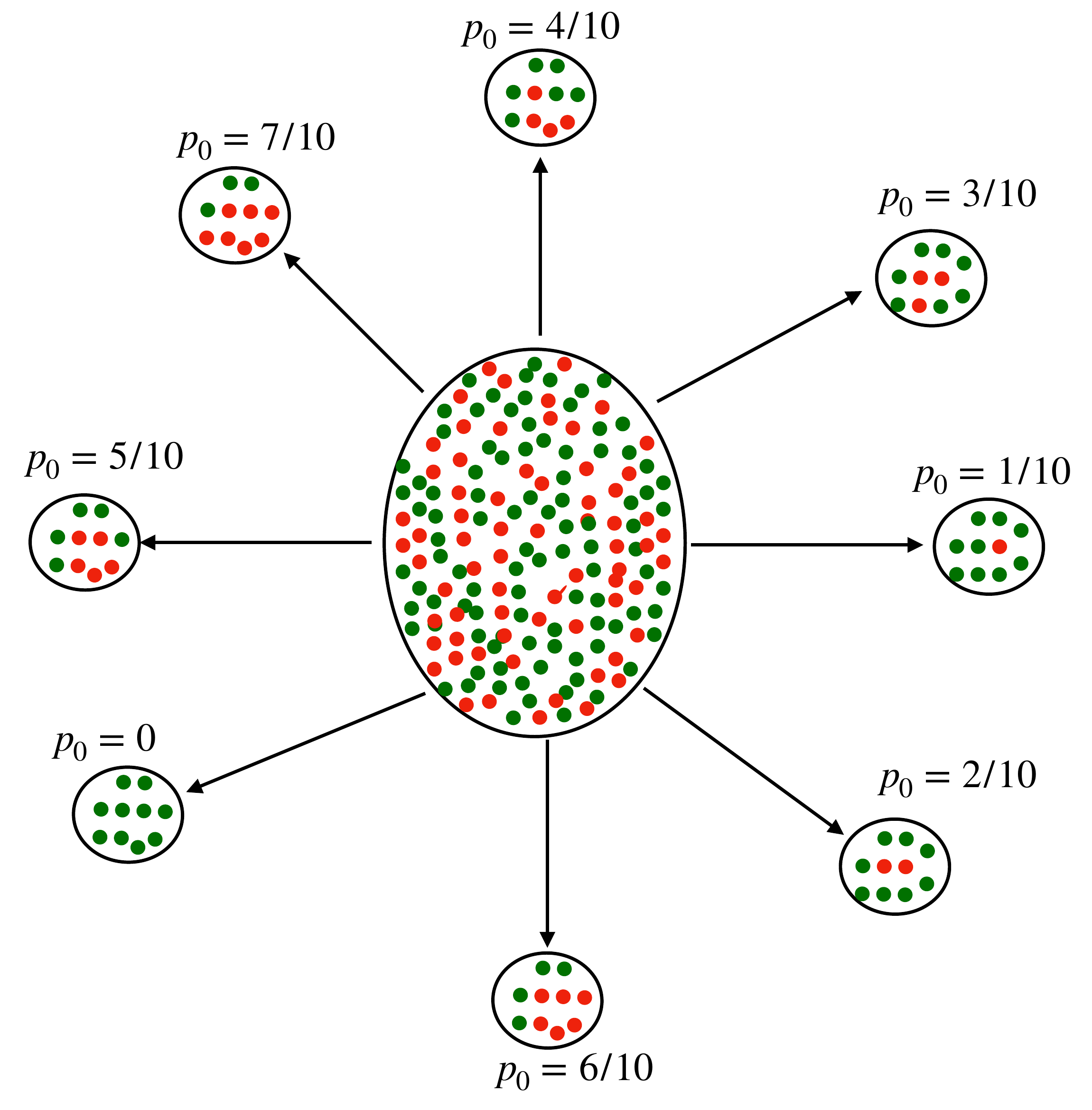}
\caption{Schematic representation of tumor containing drug-sensitive (green circles)  and drug-resistant cells (red circles) is shown at the center. For the sake of conceptual visualization,  we have shown different samples taken from the tumor, each characterized by the same number of total cells (here $N_0=10$) but different number of M-cells, and thus different values for $p_0=M/N_0$.  }
\label{fig:figp0}
\end{figure}

\noindent Within the framework of this model (Fig. \ref{fig:model}), we now address a key issue: How to characterize the {\em initial} heterogeneity (i.e. prior to the start of drug exposure) in tumor cells. 
To quantify this heterogeneity, let us consider the fraction of $M$- cells in the population. Consider the case that we isolate different sample populations from the tumor each corresponding to a fixed number ($N_0$)
of cells, see Fig. \ref{fig:figp0}.  Let $p_0=M/N_0$ denote the fraction of $M$-cells in the sample population. We propose to quantify tumor heterogeneity by considering $p_0$ as a random variable drawn from a distribution $\rho(p_0)$, characterized by its mean
$\langle p_0\rangle$  and variance $\sigma^2_{p_0}=\langle p_0^2 \rangle -\langle p_0 \rangle^2$. 
Thus, initial tumor heterogeneity is characterized not just by the presence  of drug-tolerant $M$ cells in the sample but also by variations in the number of $M$ cells from sample to sample, characterized by the distribution  $\rho(p_0)$.\\

\section{Analytical results for parameter estimation}

\noindent We now consider the stochastic process governing evolution of the tumor population upon treatment with drugs.  Upon exposure to drugs, it is a reasonable assumption that growth is inhibited, so accordingly we set 
 $k_E=k_M=0$. As discussed in the preceding section, we consider the evolution of different sample populations, each of which has a fixed initial size $N_0$ such that the fraction of $M$ cells is drawn from a distribution $\rho(p_0)$.
In this limit, the key parameters of the model are:  $k_{EM},k_{ME},\mu_E,\mu_M,\langle p_0 \rangle,\sigma^2_{p_0}$. 
In what follows, we derive analytical results that can be used to estimate model parameters by analyzing the  distribution of surviving cells upon drug exposure. \\

\noindent We note that, within our model, the evolution of each cell in the population is independent of the state of the remaining cells. 
Correspondingly, we first focus on the time evolution of a {\em single} tumor cell, which is initially either  $E$-type or $M$-type with corresponding probabilities as $1 -  p_0$ and $p_0$, 
Let us denote by $P_{E}(P_{M})$ the probability that the cell is $E$($M$)-type at time $t$, conditional on the initial probability $p_0$ for it to be $M$-type . 
The corresponding probability generating function for the single cell, conditional on the value of $p_0$ ($g(z_1,z_2,t | p_0)$
$=\sum_{\eta_E}\sum_{\eta_M}z_1^{\eta_E}z_2^{\eta_M}P(\eta_E,\eta_M,t | p_0)$), can be expressed as 
\begin{equation}{\label{ePH1}}
g(z_1,z_2,t | p_0)=1-(P_E+P_M)+P_Ez_1+P_Mz_2. 
\end{equation} 
It is straightforward to derive analytic expressions for $P_{E}(P_{M})$ and to thereby obtain an expression for $g(z_1,z_2,t | p_0)$ (Supplementary Material A). 
Now, let $G(z_1,z_2,t)$ denote the probability generating function corresponding to $P(E,M,t)$, the probability that we have ${E}$ and $ {M}$ number of sensitive ($E$-type) and resistant ($M$-type) cells in the entire population at time $t$.
Since each cell in the population (initial size $N_0$) evolves independently,
the probability generating function for the joint distribution at time $t$ (averaging over the initial choice of $p_0$) is given by
\begin{equation}{\label{ePH2}}
 G(z_1,z_2,t)=\int_{p_0=0}^{p_0=1}dp_0\rho(p_0)g(z_1,z_2,t | p_0)^{N_0},
\end{equation}
where $\rho(p_0)$ is the probability distribution function for the initial fraction of $M$-type cells ($p_0$).\\

\noindent The expression derived for the generating function, Eq. (\ref{ePH2}), can be used to derive analytic expressions for 
all the moments of the marginal distributions corresponding to $E$-type and $M$-type cells at time $t$. For example, expressions for mean number of $E$ and $M$ cells 
can be obtained using $\langle E \rangle=dG/dz_1|_{z_1=1,z_2=1}\text{and} \langle M \rangle=dG/dz_2|_{z_1=1,z_2=1}$, respectively (see
Supplementary Material A). This leads to the following expression for mean number of surviving cells at time $t$, $\langle N \rangle=\langle E+ M \rangle$:
\begin{eqnarray}{\label{eMeanN0}}
      \langle N\rangle/N_0&=&\left(\frac{\gamma_0+\alpha_0-2(\gamma_0-\mu_E+\mu_Ep_0-\mu_Mp_0)}{2\alpha_0}\right)\exp\left(-\frac{t}{2}(\gamma_0+\alpha_0)\right)\nonumber\\
     &-&\left(\frac{\gamma_0-\alpha_0-2(\gamma_0-\mu_E+\mu_Ep_0-\mu_Mp_0)}{2\alpha_0}\right)\exp\left(-\frac{t}{2}(\gamma_0-\alpha_0)\right), 
\end{eqnarray}
where
\begin{equation}{\label{eAG0}}
 \gamma_0=k_{EM}+k_{ME}+\mu_E+\mu_M,~~~~~~~
 \alpha_0=\sqrt{\gamma_0^2-4\left(k_{ME}(\mu_E-\mu_M)+(\gamma_0-\mu_M)\mu_M\right)}.
\end{equation}
It is clear from the above expression that by fitting the curve corresponding to the mean number of surviving cells as a function of time, the three parameter combinations: $\alpha_0$, $\gamma_0$, and
$\mu_E-\mu_Ep_0+\mu_Mp_0$ can be determined. \\

\noindent To extract the remaining model parameters based on time-course data, we have to turn to analytic results for the higher moments. 
For example, we can use expressions for the Fano factor ($F$) associated with total number of surviving cells, which is given by $F=\sigma^2_N/\langle N \rangle$ with $\sigma^2_N=\langle N^2 \rangle-\langle N \rangle^2$
denoting the variance in the number of surviving cells.  The expression for $\sigma^2_N$ can be obtained using 
\begin{equation}{\label{eRel}}
\sigma^2_N=\sigma^2_E+\sigma^2_M+2C_{EM},
\end{equation}
where $\sigma^2_E=\langle E^2 \rangle -\langle E \rangle^2$ and $\sigma^2_M=\langle M^2 \rangle -\langle M \rangle^2$ are variances associated with the marginal distributions for the $E$ and $M$ cells respectively, and $C_{EM}=\langle E M \rangle -\langle E \rangle \langle M \rangle$ is the correlation between numbers of $E$ and $M$ cells.
We obtain an explicit expression for the Fano factor given by (see Supplementary Material A): 
\begin{equation}{\label{efanoN0}}
  F=1-\frac{\langle N \rangle}{N_0}+\frac{N_0(N_0-1)}{\langle N \rangle}\left[\left(\frac{(\mu_E-\mu_M)}{\alpha_0}\right)
  \left(\exp\left(-\frac{t}{2}(\gamma_0-\alpha_0)\right)-\exp\left(-\frac{t}{2}
 (\gamma_0+\alpha_0)\right)\right) \right]^2\sigma^2_{p_0}.
\end{equation}
The Fano factor $F$  is a measure of deviations from the Poisson distribution, for which $F=1$. If $F<1$ or $F>1$, the distribution is
sub-Poissonian or super-Poissonian, respectively. 
Before turning our attention to approaches for parameter estimation, let us first examine the expression derived for the Fano factor.
We note that in the absence of initial variability in the fraction of $M$ cells (i.e. $\sigma^2_{p_0}=0$), the Fano
factor of surviving population is simply given by $F=1-\langle N \rangle/N_0$. As the cells are not dividing due to the exposure to drugs, the mean number of surviving cells ($\langle N \rangle$) is less than the initial population of cells ($N_0$) for $t > 0$.
Thus, in this case,  the Fano factor is \emph{always} less than one and the distribution of cell population follows
a sub-Poissonian distribution. However, given variability in the fraction of $M$-cells  in the initial population, 
the Fano factor can potentially exceed one making the distribution super-Poissonian. 
This result implies that the observation  of a Fano factor in excess of 1 in the distribution of surviving cells is an indicator of variance in the fraction of $M$-cells 
in the initial population.  Thus the measurements of the moments of surviving cell populations can provide evidence for phenotypic heterogeneity in tumor populations prior to drug treatment.\\

To gain more quantitative insight into the initial heterogeneity, we need to estimate the parameters characterizing 
the mean and variance of $\rho(p_0)$. 
 Let us rewrite Eq. (\ref{efanoN0}) in a more compact form by regrouping terms in the expression to yield the following form
\begin{eqnarray}{\label{ecF0}}
 \mathcal{F}=\left(\frac{\sigma_{p_0}(\mu_E-\mu_M)}{\alpha_0}\right)\left(\exp\left(\frac{t}{2}(\gamma_0-\alpha_0)\right)-\exp\left(\frac{t}{2}
 (\gamma_0+\alpha_0)\right)\right),
\end{eqnarray}
with 
\begin{equation}
 \mathcal{F}=\exp\left[\frac{\ln \left(\left(F-1+\frac{\langle N \rangle}{N_0}\right)\frac{\langle N \rangle}{N_0(N_0-1)}\right)}{2}\right].
\end{equation}

\noindent Using the expressions for the mean and Fano factor of the surviving population as functions of time, Eqs. (\ref{eMeanN0}) and (\ref{ecF0}), we can estimate four of the parameter
combinations, namely, $\alpha_0$, $\gamma_0$, $\mu_E-\mu_Ep_0+\mu_Mp_0$, and $(\sigma_{p_0}(\mu_E-\mu_M))/\alpha_0$. Correspondingly, we need additional experiments to determine the entire set of 6 model parameters. 
As we now show, a set of measurements that accomplish this can be obtained 
by starting from different initial conditions. \\

\noindent The proposed protocol is motivated by that fact that experimental techniques such as fluorescence-activated cell sorting (FACS) can be 
used to prepare the samples in specified initial states. 
With this in mind, we begin from an initial condition where all cells are $E$-type i.e. $p_0=0, \sigma^2_{p_0}=0$. 
Using the derived results, we can determine the parameters 
$\mu_E$, $\alpha_0$ and $\gamma_0$. We next consider the initial condition to be all $M$-type cells i.e.  $p_0=1, \sigma^2_{p_0}=0$.
Analysis of the corresponding time-course measurements of the number of surviving cells can now be used to estimate the parameter $\mu_M$. 
Having estimated values of $\alpha_0$, $\gamma_0$, $\mu_E$, and $\mu_M$, we can now find the 
switching rates, $k_{EM}$ and $k_{ME}$, from Eq.(\ref{eAG0}). Finally, using the expressions for the mean number of surviving cells and corresponding Fano factors for arbitrary $p_0$, we can
get explicit expressions for the probability $p_0$ and variance $\sigma^2_{p_0}$ (Supplementary Material A): 
\begin{eqnarray}
 p_0&=& \frac{\langle N \rangle-\langle N \rangle_0}{\langle N \rangle_1-\langle N \rangle_0},\nonumber\\
 \sigma^2_{p_0}&=&\frac{N_0\langle N \rangle}{(\langle N \rangle_1-\langle N \rangle_0)^2(N_0-1)}\left[F-1+\frac{\langle N \rangle}{N_0}\right].
\end{eqnarray}
The above results are expressed in terms of experimentally measurable quantities, involving mean  values  $\langle N \rangle_0$ (for $p_0=0$) and  $\langle N \rangle_1$ (for $p_0=1$) and the Fano-factor of the total
surviving population, and thus can be used to estimate the population heterogeneity at the start of drug exposure based on time-course measurements of the surviving population size. \\

\section{Modeling generation of tumor heterogeneity}
The analysis in the preceding section holds regardless of the source of initial heterogeneity in tumor populations. In this section, we explore how the model introduced for 
tumor cell dynamics can also be used to analyze a potential mechanism for generation of tumor heterogeneity. We note that the proposed model in Fig. 2 can be seen as a generalized version of  the celebrated 
 Luria-Delbr{\"u}ck (LD) model with the important addition that, in the present case, the transition between the two phenotypes is reversible (as opposed to the  Luria-Delbr{\"u}ck case).
However, while the LD model can be solved exactly \cite{antal2011exact}, the exact analytical solution of the reversible model in Fig. 2 is not known, to the best of our knowledge. 
Nevertheless, as we show below, exact expressions for the mean and variance of the number of $E$-type and $M$-type cells can be obtained and used to 
characterize heterogeneity in tumor cell populations. We can use the master equation, Eq.(\ref{eMaster}), to derive expressions 
for the mean number of $E$ and $M$ cells at any time $t$ (Supplementary Material B). Using these
expressions, the mean number of surviving cells $\langle N\rangle=$ $\langle E\rangle+\langle M \rangle$ is given by:
\begin{eqnarray}{\label{eqN}}
 \langle N \rangle&=&\left(\frac{E_0(\alpha-\gamma-2k_E^f)+M_0(\alpha-\gamma-2k_M^f)}{2\alpha}\right)\exp\left(-\frac{t}{2}(\gamma+\alpha)\right)
 \nonumber\\&+&\left(\frac{E_0(\alpha+\gamma+2k_E^f)+M_0(\alpha+\gamma+2k_M^f)}{2\alpha}\right)\exp\left(-\frac{t}{2}(\gamma-\alpha)\right),
\end{eqnarray}
where
\begin{eqnarray}{\label{eqgmal}}
 \gamma&=&k_{EM}+k_{ME}-k_E^f-k_M^f, ~~~~~~~~
 \alpha=\sqrt{\gamma^2+4\left(k_{ME}(k_E^f-k_M^f)+(\gamma+k_M^f)k_M^f\right)},\nonumber\\
\end{eqnarray} 
with  
$k_E^f=k_E-\mu_E$  and  $k_M^f=k_M-\mu_M$
representing the effective birth rates for $E$-type and $M$-type cells respectively, while
$E_0$ and $M_0$ are the initial numbers of $E$ and $M$ cells at $t=0$. \\

The results show that the mean number of surviving cells at any time $t$ is characterized by six parameters: initial number of $E$ and $M$ cells ($E_0,M_0$), two effective birth 
rates ($k_E^f$, $k_M^f$) and two switching rates ($k_{EM}, k_{ME}$).
Given that the initial population can be chosen in a controlled manner, we can use the results for the mean population size to determine some of the model parameters.
Specifically, we can set $M_0=0$ as the initial condition and fitting the data to obtain  the coefficient of exponential terms in Eq. (\ref{eqN}) will yield $k_E^f$. Next, we can set  $E_0=0$ and Eq. (\ref{eqN}) 
will allow us to extract $k_M^f$. Once we estimate $k_E^f$ and $k_M^f$, we can extract the switching rates using the estimated values of $\gamma$ and $\alpha$,
using Eq. (\ref{eqgmal}). That is, the proposed procedure allows us to estimate the parameter combinations, $k_E^f$ and $k_M^f$ as well as the parameters $k_{EM}$ and $k_{ME}$. \\

In order to estimate the remaining model parameters, we need
to consider the higher moments. While obtaining analytical expressions for the full probability distribution is still
an open problem, higher moments can be calculated in a straightforward manner. For example, using Eq. (\ref{eMaster}) the evolution equation for 
$\langle E^2 \rangle=\sum E^2 P(E,M,t)$, $\langle M^2 \rangle=\sum M^2 P(E,M,t)$ and $\langle EM \rangle=\sum EM P(E,M,t)$ is given by
\begin{eqnarray}{\label{e2nd}}
 \frac{\partial \langle E^2\rangle}{\partial t}&=&(k_E+\mu_E+k_{EM}) \langle E \rangle+k_{ME}\langle M \rangle+2(k_E-\mu_E-k_{EM})\langle E^2\rangle+2k_{ME}\langle ME \rangle,\nonumber\\
\frac{\partial \langle M^2\rangle}{\partial t}&=&(k_M+\mu_M+k_{ME})\langle M \rangle+k_{EM}\langle E \rangle+2(k_M-\mu_M-k_{ME})\langle M^2\rangle+2k_{EM}\langle ME \rangle,\nonumber\\
\frac{\partial \langle ME \rangle}{\partial t}&=&-k_{ME}\langle M\rangle-k_{EM}\langle E \rangle+(k_E+k_M-\mu_E-\mu_M-k_{ME}-k_{EM})\langle ME \rangle +k_{ME}\langle M^2\rangle
+k_{EM}\langle E^2\rangle.
\end{eqnarray}
The above set of equations can be solved to get explicit expressions for $\langle E^2 \rangle$, $\langle M^2 \rangle$ and $\langle EM \rangle$ at any time $t$ (Supplementary Material B), which can be used to get the variance ($\sigma^2_N$) in the total number of surviving cells by using Eq. (\ref{eRel}). 
The analytic expression for the variance, in combination with the expression for mean number of surviving cells, can be used to extract all model parameters. 
Furthermore, the extracted parameters can then be compared with the parameters derived based on tumor cell dynamics {\em after} exposure to drugs. The comparisons can provide insight into the relative roles of adaptation and selection in driving tumor 
heterogeneity. The scenario wherein the switching parameters $k_{EM}$ and $k_{ME}$ are effectively unchanged upon exposure to drugs, whereas $\mu_E$ and $\mu_M$ increase favors selection as the dominant driver of tumor heterogeneity. However, significant changes in 
the switching rates  $k_{EM}$ and $k_{ME}$ are indicative of a role for adaptation as well in the generation of tumor heterogeneity. \\

\section{Characterizing the distribution of the fraction of resistant cells}
\begin{figure}[h]
\centering
\includegraphics[angle=90,width=14cm]{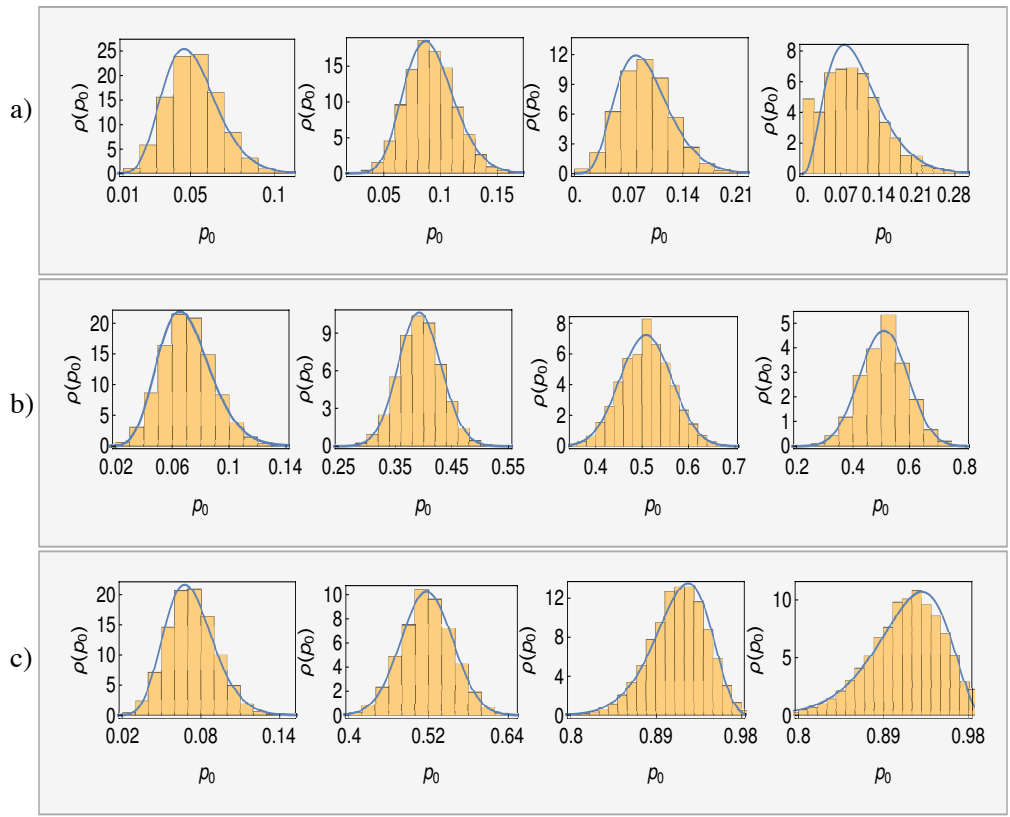}
\caption{Simulation results for the distributions of fraction of M-cells $\rho(p_0)$ are shown as histograms and the continuous solid lines represent fits by Beta distributions. Top (a), middle (b) and lower(c) panels corresponds to $k_{EM}/k_{ME}$=0.1, 1 and 10 respectively. In each panel, distributions are shown at various time points, $t=0.1,1,10,20$ from left to right. Other parameters are: $k_E=0.2,k_M=0.1,\mu_E=0.3,\mu_M=0.15$.} 
\label{fig:beta}
\end{figure}

\noindent The results derived for the moments can also be used to characterize the probability distribution $\rho(p_0)$ for the fraction of $M$-type cells. Recall that we must have $0 \le p_0 \le 1$ and furthermore the first two moments of $\rho(p_0)$ can be obtained using the procedure 
outlined in the previous section. Thus a natural choice to characterize the distribution $\rho(p_0)$ is to take it to be 
the Beta distribution with the mean and variance as determined by measurements. Such a distribution is expressed in terms of two exponents ($\alpha$ and $\beta$) as
\begin{equation}
\rho(p_0)=\frac{\Gamma(\alpha+\beta)}{\Gamma(\alpha)\Gamma(\beta)}p_0^{\alpha-1}(1-p_0)^{\beta-1}
\end{equation}
with mean $\langle p_0 \rangle=\frac{\alpha}{\alpha+\beta} \text{
and variance}~~ \sigma^2_{p_0}=\frac{\alpha\beta}{(\alpha+\beta)^2(\alpha+\beta+1)}.$
The two parameters of the Beta distribution can be estimated using the experimentally determined mean and variance. Explicitly, these
are given by
\begin{equation}
\alpha=\frac{\langle p_0\rangle\left(\langle p_0\rangle(1-\langle p_0\rangle)-\sigma^2_{p_0}\right)}{\sigma^2_{p_0}}, ~~~\beta=\frac{(\langle p_0 \rangle-1)\left[\langle p_0 \rangle^2-\langle p_0 \rangle+\sigma^2_{p_0}\right]}{\sigma^2_{p_0}}.
\end{equation}
To test this approach for characterizing the initial heterogeneity, we compare the Beta distribution 
with the results obtained from stochastic simulations of the model. Specifically, we carried out stochastic simulations using the Gillespie algorithm \cite{gillespie1977exact} for the model in Fig.2 starting with 200 sensitive E-type cells and no resistant M-type cells.
The empirically determined distribution for the fraction of $M$-type cells ($\rho(p_0)$) is then compared to the Beta distribution with the same mean and variance
as the empirical distribution. The results obtained are shown in Fig.~\ref{fig:beta}, which indicate that the Beta distribution is an excellent approximation for the range of parameters explored.
\section{Discussion}
\noindent To summarize, we have studied a coarse-grained stochastic model to quantify phenotypic heterogeneity in a population of cancer cells. Motivated by the experimental observation that both chemoresistance and TGF-$\beta$ induced EMT lead to similar outcomes, the model assumes that a cell has two phenotypes corresponding to whether it is drug-sensitive or drug-resistant. Importantly, the model is also consistent with  epigenetic mechanisms for generating phenotypic heterogeneity in cancer, given that it allows reversible phenotypic switching between sensitive and resistant cells. \\

\noindent For the model considered, we have derived analytic results, both in the presence and absence of chemotherapeutic agents,  which provide insights into the role of phenotypic switching in generating 
population heterogeneity. One of the issues that we address through these results focuses on quantifying initial heterogeneity in the fraction of resistant cells, characterized by mean fraction of resistant cells $\langle p_0 \rangle $ in a tumor and its variance $\sigma^2_{p_0}$. For this, we propose a protocol that can be used to estimate the model parameters based on measurements of mean and variance of the surviving population of tumor cells. 
Furthermore, our analysis also leads to a condition, in terms of experimentally measurable quantities, whose value serves as an indicator for the presence of initial heterogeneity in the fraction of resistant cells. \\

 \noindent While the proposed method allows us to estimate the mean and variance of the fraction of resistant cells prior to therapy, obtaining an exact analytical form for the entire distribution appears to be challenging. However, our simulation results 
 suggest that this distribution is well approximated by the Beta distribution, which can be characterized by using the mean and variance of the surviving population. Besides characterizing initial heterogeneity in the cancer cell population,  the estimated model parameters can also be useful in analyzing the complex roles of adaptation and selection in the acquisition of chemoresistance.  Furthermore, the results obtained provide exact analytical expressions characterizing the distribution of of tumor cell population under treatment by drugs. Going forward we envision further model development in dialog with experiments that longitudinally monitor phenotypic changes in time lapse microscopy studies, either by quantitative analysis of morphometric parameters or implementing fluorescent reporters of EMT which have been recently developed \cite{toneff2016z}.  These results can serve as important inputs to future work focusing on evaluation of the hypothesis that model-informed design of treatment schedule and dose parameters may reduce the emergence of chemoresistance.\\
 

\noindent{\bf Acknowledgments: }The authors gratefully acknowledge funding support from the NIH through grants 3U54CA156734-05S3 (as part of the  UMass Boston/Dana Farber-Harvard Cancer Center U54 partnership)
and from the National Cancer Institute R00CA155045 ( PI: JPC), and a Sanofi Genzyme doctoral fellowship which supported Gwendolyn Cramer. We would also like to acknowledge funding support from the UMass Boston Healey award.\\

\noindent{\bf Author contributions:}
NK, BS, JC and RVK designed research; NK, SZD and RK carried out theoretical/ computational aspects of research, GC carried out experimental aspects of research; NK, BS, JC and RVK wrote the paper and all authors reviewed the paper. 

\bibliography{healey}

\begin{thebibliography}{53}
\expandafter\ifx\csname natexlab\endcsname\relax\def\natexlab#1{#1}\fi
\expandafter\ifx\csname bibnamefont\endcsname\relax
  \def\bibnamefont#1{#1}\fi
\expandafter\ifx\csname bibfnamefont\endcsname\relax
  \def\bibfnamefont#1{#1}\fi
\expandafter\ifx\csname citenamefont\endcsname\relax
  \def\citenamefont#1{#1}\fi
\expandafter\ifx\csname url\endcsname\relax
  \def\url#1{\texttt{#1}}\fi
\expandafter\ifx\csname urlprefix\endcsname\relax\def\urlprefix{URL }\fi
\providecommand{\bibinfo}[2]{#2}
\providecommand{\eprint}[2][]{\url{#2}}

\bibitem[{\citenamefont{Housman et~al.}(2014)\citenamefont{Housman, Byler,
  Heerboth, Lapinska, Longacre, Snyder, and Sarkar}}]{housman2014drug}
\bibinfo{author}{\bibfnamefont{G.}~\bibnamefont{Housman}},
  \bibinfo{author}{\bibfnamefont{S.}~\bibnamefont{Byler}},
  \bibinfo{author}{\bibfnamefont{S.}~\bibnamefont{Heerboth}},
  \bibinfo{author}{\bibfnamefont{K.}~\bibnamefont{Lapinska}},
  \bibinfo{author}{\bibfnamefont{M.}~\bibnamefont{Longacre}},
  \bibinfo{author}{\bibfnamefont{N.}~\bibnamefont{Snyder}}, \bibnamefont{and}
  \bibinfo{author}{\bibfnamefont{S.}~\bibnamefont{Sarkar}},
  \bibinfo{journal}{Cancers} \textbf{\bibinfo{volume}{6}},
  \bibinfo{pages}{1769} (\bibinfo{year}{2014}).

\bibitem[{\citenamefont{Gottesman}(2002)}]{gottesman2002mechanisms}
\bibinfo{author}{\bibfnamefont{M.~M.} \bibnamefont{Gottesman}},
  \bibinfo{journal}{Annual review of medicine} \textbf{\bibinfo{volume}{53}},
  \bibinfo{pages}{615} (\bibinfo{year}{2002}).

\bibitem[{\citenamefont{Chisholm et~al.}(2016)\citenamefont{Chisholm, Lorenzi,
  and Clairambault}}]{chisholm2016cell}
\bibinfo{author}{\bibfnamefont{R.~H.} \bibnamefont{Chisholm}},
  \bibinfo{author}{\bibfnamefont{T.}~\bibnamefont{Lorenzi}}, \bibnamefont{and}
  \bibinfo{author}{\bibfnamefont{J.}~\bibnamefont{Clairambault}},
  \bibinfo{journal}{Biochimica et Biophysica Acta (BBA)-General Subjects}
  \textbf{\bibinfo{volume}{1860}}, \bibinfo{pages}{2627}
  (\bibinfo{year}{2016}).

\bibitem[{\citenamefont{Bozic and Nowak}(2017)}]{bozic2017resisting}
\bibinfo{author}{\bibfnamefont{I.}~\bibnamefont{Bozic}} \bibnamefont{and}
  \bibinfo{author}{\bibfnamefont{M.~A.} \bibnamefont{Nowak}},
  \bibinfo{journal}{Annual Review of Cancer Biology}
  \textbf{\bibinfo{volume}{1}}, \bibinfo{pages}{203} (\bibinfo{year}{2017}).

\bibitem[{\citenamefont{Pogrebniak and
  Curtis}(2018)}]{pogrebniak2018harnessing}
\bibinfo{author}{\bibfnamefont{K.~L.} \bibnamefont{Pogrebniak}}
  \bibnamefont{and} \bibinfo{author}{\bibfnamefont{C.~N.}
  \bibnamefont{Curtis}}, \bibinfo{journal}{Trends in Genetics}
  (\bibinfo{year}{2018}).

\bibitem[{\citenamefont{Nikolaou et~al.}(2018)\citenamefont{Nikolaou,
  Pavlopoulou, Georgakilas, and Kyrodimos}}]{nikolaou2018challenge}
\bibinfo{author}{\bibfnamefont{M.}~\bibnamefont{Nikolaou}},
  \bibinfo{author}{\bibfnamefont{A.}~\bibnamefont{Pavlopoulou}},
  \bibinfo{author}{\bibfnamefont{A.~G.} \bibnamefont{Georgakilas}},
  \bibnamefont{and}
  \bibinfo{author}{\bibfnamefont{E.}~\bibnamefont{Kyrodimos}},
  \bibinfo{journal}{Clinical \& Experimental Metastasis} pp.
  \bibinfo{pages}{1--10} (\bibinfo{year}{2018}).

\bibitem[{\citenamefont{Salgia and Kulkarni}(2018)}]{salgia2018genetic}
\bibinfo{author}{\bibfnamefont{R.}~\bibnamefont{Salgia}} \bibnamefont{and}
  \bibinfo{author}{\bibfnamefont{P.}~\bibnamefont{Kulkarni}},
  \bibinfo{journal}{Trends in cancer}  (\bibinfo{year}{2018}).

\bibitem[{\citenamefont{Dagogo-Jack and Shaw}(2018)}]{dagogo2018tumour}
\bibinfo{author}{\bibfnamefont{I.}~\bibnamefont{Dagogo-Jack}} \bibnamefont{and}
  \bibinfo{author}{\bibfnamefont{A.~T.} \bibnamefont{Shaw}},
  \bibinfo{journal}{Nature reviews Clinical oncology}
  \textbf{\bibinfo{volume}{15}}, \bibinfo{pages}{81} (\bibinfo{year}{2018}).

\bibitem[{\citenamefont{Zhou et~al.}(2009)\citenamefont{Zhou, Zhang, Damelin,
  Geles, Grindley, and Dirks}}]{zhou2009tumour}
\bibinfo{author}{\bibfnamefont{B.-B.~S.} \bibnamefont{Zhou}},
  \bibinfo{author}{\bibfnamefont{H.}~\bibnamefont{Zhang}},
  \bibinfo{author}{\bibfnamefont{M.}~\bibnamefont{Damelin}},
  \bibinfo{author}{\bibfnamefont{K.~G.} \bibnamefont{Geles}},
  \bibinfo{author}{\bibfnamefont{J.~C.} \bibnamefont{Grindley}},
  \bibnamefont{and} \bibinfo{author}{\bibfnamefont{P.~B.} \bibnamefont{Dirks}},
  \bibinfo{journal}{Nature reviews Drug discovery}
  \textbf{\bibinfo{volume}{8}}, \bibinfo{pages}{806} (\bibinfo{year}{2009}).

\bibitem[{\citenamefont{Zahreddine and
  Borden}(2013)}]{zahreddine2013mechanisms}
\bibinfo{author}{\bibfnamefont{H.}~\bibnamefont{Zahreddine}} \bibnamefont{and}
  \bibinfo{author}{\bibfnamefont{K.}~\bibnamefont{Borden}},
  \bibinfo{journal}{Frontiers in pharmacology} \textbf{\bibinfo{volume}{4}},
  \bibinfo{pages}{28} (\bibinfo{year}{2013}).

\bibitem[{\citenamefont{Holohan et~al.}(2013)\citenamefont{Holohan,
  Van~Schaeybroeck, Longley, and Johnston}}]{holohan2013cancer}
\bibinfo{author}{\bibfnamefont{C.}~\bibnamefont{Holohan}},
  \bibinfo{author}{\bibfnamefont{S.}~\bibnamefont{Van~Schaeybroeck}},
  \bibinfo{author}{\bibfnamefont{D.~B.} \bibnamefont{Longley}},
  \bibnamefont{and} \bibinfo{author}{\bibfnamefont{P.~G.}
  \bibnamefont{Johnston}}, \bibinfo{journal}{Nature Reviews Cancer}
  \textbf{\bibinfo{volume}{13}}, \bibinfo{pages}{714} (\bibinfo{year}{2013}).

\bibitem[{\citenamefont{Garraway and
  J{\"a}nne}(2012)}]{garraway2012circumventing}
\bibinfo{author}{\bibfnamefont{L.~A.} \bibnamefont{Garraway}} \bibnamefont{and}
  \bibinfo{author}{\bibfnamefont{P.~A.} \bibnamefont{J{\"a}nne}},
  \bibinfo{journal}{Cancer discovery} \textbf{\bibinfo{volume}{2}},
  \bibinfo{pages}{214} (\bibinfo{year}{2012}).

\bibitem[{\citenamefont{Shaffer et~al.}(2017)\citenamefont{Shaffer, Dunagin,
  Torborg, Torre, Emert, Krepler, Beqiri, Sproesser, Brafford, Xiao
  et~al.}}]{shaffer2017rare}
\bibinfo{author}{\bibfnamefont{S.~M.} \bibnamefont{Shaffer}},
  \bibinfo{author}{\bibfnamefont{M.~C.} \bibnamefont{Dunagin}},
  \bibinfo{author}{\bibfnamefont{S.~R.} \bibnamefont{Torborg}},
  \bibinfo{author}{\bibfnamefont{E.~A.} \bibnamefont{Torre}},
  \bibinfo{author}{\bibfnamefont{B.}~\bibnamefont{Emert}},
  \bibinfo{author}{\bibfnamefont{C.}~\bibnamefont{Krepler}},
  \bibinfo{author}{\bibfnamefont{M.}~\bibnamefont{Beqiri}},
  \bibinfo{author}{\bibfnamefont{K.}~\bibnamefont{Sproesser}},
  \bibinfo{author}{\bibfnamefont{P.~A.} \bibnamefont{Brafford}},
  \bibinfo{author}{\bibfnamefont{M.}~\bibnamefont{Xiao}}, \bibnamefont{et~al.},
  \bibinfo{journal}{Nature} \textbf{\bibinfo{volume}{546}},
  \bibinfo{pages}{431} (\bibinfo{year}{2017}).

\bibitem[{\citenamefont{Chabner and Roberts}(2005)}]{chabner2005chemotherapy}
\bibinfo{author}{\bibfnamefont{B.~A.} \bibnamefont{Chabner}} \bibnamefont{and}
  \bibinfo{author}{\bibfnamefont{T.~G.} \bibnamefont{Roberts}},
  \bibinfo{journal}{Nature Reviews Cancer} \textbf{\bibinfo{volume}{5}},
  \bibinfo{pages}{65} (\bibinfo{year}{2005}).

\bibitem[{\citenamefont{Gatenby and Brown}(2018)}]{gatenby2018evolution}
\bibinfo{author}{\bibfnamefont{R.}~\bibnamefont{Gatenby}} \bibnamefont{and}
  \bibinfo{author}{\bibfnamefont{J.}~\bibnamefont{Brown}},
  \bibinfo{journal}{Cold Spring Harbor perspectives in medicine}
  \textbf{\bibinfo{volume}{8}}, \bibinfo{pages}{a033415}
  (\bibinfo{year}{2018}).

\bibitem[{\citenamefont{Gallaher et~al.}(2018)\citenamefont{Gallaher,
  Enriquez-Navas, Luddy, Gatenby, and Anderson}}]{gallaher2018spatial}
\bibinfo{author}{\bibfnamefont{J.}~\bibnamefont{Gallaher}},
  \bibinfo{author}{\bibfnamefont{P.~M.} \bibnamefont{Enriquez-Navas}},
  \bibinfo{author}{\bibfnamefont{K.~A.} \bibnamefont{Luddy}},
  \bibinfo{author}{\bibfnamefont{R.~A.} \bibnamefont{Gatenby}},
  \bibnamefont{and} \bibinfo{author}{\bibfnamefont{A.~R.}
  \bibnamefont{Anderson}}, \bibinfo{journal}{Cancer research} pp.
  \bibinfo{pages}{canres--2649} (\bibinfo{year}{2018}).

\bibitem[{\citenamefont{Castorina et~al.}(2009)\citenamefont{Castorina,
  Carc{\`o}, Guiot, and Deisboeck}}]{castorina2009tumor}
\bibinfo{author}{\bibfnamefont{P.}~\bibnamefont{Castorina}},
  \bibinfo{author}{\bibfnamefont{D.}~\bibnamefont{Carc{\`o}}},
  \bibinfo{author}{\bibfnamefont{C.}~\bibnamefont{Guiot}}, \bibnamefont{and}
  \bibinfo{author}{\bibfnamefont{T.~S.} \bibnamefont{Deisboeck}},
  \bibinfo{journal}{Cancer research} \textbf{\bibinfo{volume}{69}},
  \bibinfo{pages}{8507} (\bibinfo{year}{2009}).

\bibitem[{\citenamefont{Pardal et~al.}(2003)\citenamefont{Pardal, Clarke, and
  Morrison}}]{pardal2003applying}
\bibinfo{author}{\bibfnamefont{R.}~\bibnamefont{Pardal}},
  \bibinfo{author}{\bibfnamefont{M.~F.} \bibnamefont{Clarke}},
  \bibnamefont{and} \bibinfo{author}{\bibfnamefont{S.~J.}
  \bibnamefont{Morrison}}, \bibinfo{journal}{Nature Reviews Cancer}
  \textbf{\bibinfo{volume}{3}}, \bibinfo{pages}{895} (\bibinfo{year}{2003}).

\bibitem[{\citenamefont{Meacham and Morrison}(2013)}]{meacham2013tumour}
\bibinfo{author}{\bibfnamefont{C.~E.} \bibnamefont{Meacham}} \bibnamefont{and}
  \bibinfo{author}{\bibfnamefont{S.~J.} \bibnamefont{Morrison}},
  \bibinfo{journal}{Nature} \textbf{\bibinfo{volume}{501}},
  \bibinfo{pages}{328} (\bibinfo{year}{2013}).

\bibitem[{\citenamefont{Marusyk et~al.}(2012)\citenamefont{Marusyk, Almendro,
  and Polyak}}]{marusyk2012intra}
\bibinfo{author}{\bibfnamefont{A.}~\bibnamefont{Marusyk}},
  \bibinfo{author}{\bibfnamefont{V.}~\bibnamefont{Almendro}}, \bibnamefont{and}
  \bibinfo{author}{\bibfnamefont{K.}~\bibnamefont{Polyak}},
  \bibinfo{journal}{Nature Reviews Cancer} \textbf{\bibinfo{volume}{12}},
  \bibinfo{pages}{323} (\bibinfo{year}{2012}).

\bibitem[{\citenamefont{Gupta et~al.}(2009)\citenamefont{Gupta, Chaffer, and
  Weinberg}}]{gupta2009cancer}
\bibinfo{author}{\bibfnamefont{P.~B.} \bibnamefont{Gupta}},
  \bibinfo{author}{\bibfnamefont{C.~L.} \bibnamefont{Chaffer}},
  \bibnamefont{and} \bibinfo{author}{\bibfnamefont{R.~A.}
  \bibnamefont{Weinberg}}, \bibinfo{journal}{Nature medicine}
  \textbf{\bibinfo{volume}{15}}, \bibinfo{pages}{1010} (\bibinfo{year}{2009}).

\bibitem[{\citenamefont{Zhou et~al.}(2014)\citenamefont{Zhou, Pisco, Qian, and
  Huang}}]{zhou2014nonequilibrium}
\bibinfo{author}{\bibfnamefont{J.~X.} \bibnamefont{Zhou}},
  \bibinfo{author}{\bibfnamefont{A.~O.} \bibnamefont{Pisco}},
  \bibinfo{author}{\bibfnamefont{H.}~\bibnamefont{Qian}}, \bibnamefont{and}
  \bibinfo{author}{\bibfnamefont{S.}~\bibnamefont{Huang}},
  \bibinfo{journal}{PloS one} \textbf{\bibinfo{volume}{9}},
  \bibinfo{pages}{e110714} (\bibinfo{year}{2014}).

\bibitem[{\citenamefont{Nowell}(1976)}]{nowell1976clonal}
\bibinfo{author}{\bibfnamefont{P.~C.} \bibnamefont{Nowell}},
  \bibinfo{journal}{Science} \textbf{\bibinfo{volume}{194}},
  \bibinfo{pages}{23} (\bibinfo{year}{1976}).

\bibitem[{\citenamefont{Sottoriva et~al.}(2013)\citenamefont{Sottoriva,
  Spiteri, Piccirillo, Touloumis, Collins, Marioni, Curtis, Watts, and
  Tavar{\'e}}}]{sottoriva2013intratumor}
\bibinfo{author}{\bibfnamefont{A.}~\bibnamefont{Sottoriva}},
  \bibinfo{author}{\bibfnamefont{I.}~\bibnamefont{Spiteri}},
  \bibinfo{author}{\bibfnamefont{S.~G.} \bibnamefont{Piccirillo}},
  \bibinfo{author}{\bibfnamefont{A.}~\bibnamefont{Touloumis}},
  \bibinfo{author}{\bibfnamefont{V.~P.} \bibnamefont{Collins}},
  \bibinfo{author}{\bibfnamefont{J.~C.} \bibnamefont{Marioni}},
  \bibinfo{author}{\bibfnamefont{C.}~\bibnamefont{Curtis}},
  \bibinfo{author}{\bibfnamefont{C.}~\bibnamefont{Watts}}, \bibnamefont{and}
  \bibinfo{author}{\bibfnamefont{S.}~\bibnamefont{Tavar{\'e}}},
  \bibinfo{journal}{Proceedings of the National Academy of Sciences}
  \textbf{\bibinfo{volume}{110}}, \bibinfo{pages}{4009} (\bibinfo{year}{2013}).

\bibitem[{\citenamefont{Burrell et~al.}(2013)\citenamefont{Burrell, McGranahan,
  Bartek, and Swanton}}]{burrell2013causes}
\bibinfo{author}{\bibfnamefont{R.~A.} \bibnamefont{Burrell}},
  \bibinfo{author}{\bibfnamefont{N.}~\bibnamefont{McGranahan}},
  \bibinfo{author}{\bibfnamefont{J.}~\bibnamefont{Bartek}}, \bibnamefont{and}
  \bibinfo{author}{\bibfnamefont{C.}~\bibnamefont{Swanton}},
  \bibinfo{journal}{Nature} \textbf{\bibinfo{volume}{501}},
  \bibinfo{pages}{338} (\bibinfo{year}{2013}).

\bibitem[{\citenamefont{Pisco et~al.}(2013)\citenamefont{Pisco, Brock, Zhou,
  Moor, Mojtahedi, Jackson, and Huang}}]{pisco2013non}
\bibinfo{author}{\bibfnamefont{A.~O.} \bibnamefont{Pisco}},
  \bibinfo{author}{\bibfnamefont{A.}~\bibnamefont{Brock}},
  \bibinfo{author}{\bibfnamefont{J.}~\bibnamefont{Zhou}},
  \bibinfo{author}{\bibfnamefont{A.}~\bibnamefont{Moor}},
  \bibinfo{author}{\bibfnamefont{M.}~\bibnamefont{Mojtahedi}},
  \bibinfo{author}{\bibfnamefont{D.}~\bibnamefont{Jackson}}, \bibnamefont{and}
  \bibinfo{author}{\bibfnamefont{S.}~\bibnamefont{Huang}},
  \bibinfo{journal}{Nature communications} \textbf{\bibinfo{volume}{4}},
  \bibinfo{pages}{2467} (\bibinfo{year}{2013}).

\bibitem[{\citenamefont{Pisco and Huang}(2015)}]{pisco2015non}
\bibinfo{author}{\bibfnamefont{A.~O.} \bibnamefont{Pisco}} \bibnamefont{and}
  \bibinfo{author}{\bibfnamefont{S.}~\bibnamefont{Huang}},
  \bibinfo{journal}{British journal of cancer} \textbf{\bibinfo{volume}{112}},
  \bibinfo{pages}{1725} (\bibinfo{year}{2015}).

\bibitem[{\citenamefont{Brown et~al.}(2014)\citenamefont{Brown, Curry, Magnani,
  Wilhelm-Benartzi, and Borley}}]{brown2014poised}
\bibinfo{author}{\bibfnamefont{R.}~\bibnamefont{Brown}},
  \bibinfo{author}{\bibfnamefont{E.}~\bibnamefont{Curry}},
  \bibinfo{author}{\bibfnamefont{L.}~\bibnamefont{Magnani}},
  \bibinfo{author}{\bibfnamefont{C.~S.} \bibnamefont{Wilhelm-Benartzi}},
  \bibnamefont{and} \bibinfo{author}{\bibfnamefont{J.}~\bibnamefont{Borley}},
  \bibinfo{journal}{Nature Reviews Cancer} \textbf{\bibinfo{volume}{14}},
  \bibinfo{pages}{747} (\bibinfo{year}{2014}).

\bibitem[{\citenamefont{Su et~al.}(2017)\citenamefont{Su, Wei, Robert, Xue,
  Tsoi, Garcia-Diaz, Moreno, Kim, Ng, Lee et~al.}}]{su2017single}
\bibinfo{author}{\bibfnamefont{Y.}~\bibnamefont{Su}},
  \bibinfo{author}{\bibfnamefont{W.}~\bibnamefont{Wei}},
  \bibinfo{author}{\bibfnamefont{L.}~\bibnamefont{Robert}},
  \bibinfo{author}{\bibfnamefont{M.}~\bibnamefont{Xue}},
  \bibinfo{author}{\bibfnamefont{J.}~\bibnamefont{Tsoi}},
  \bibinfo{author}{\bibfnamefont{A.}~\bibnamefont{Garcia-Diaz}},
  \bibinfo{author}{\bibfnamefont{B.~H.} \bibnamefont{Moreno}},
  \bibinfo{author}{\bibfnamefont{J.}~\bibnamefont{Kim}},
  \bibinfo{author}{\bibfnamefont{R.~H.} \bibnamefont{Ng}},
  \bibinfo{author}{\bibfnamefont{J.~W.} \bibnamefont{Lee}},
  \bibnamefont{et~al.}, \bibinfo{journal}{Proceedings of the National Academy
  of Sciences} p. \bibinfo{pages}{201712064} (\bibinfo{year}{2017}).

\bibitem[{\citenamefont{Inde and Dixon}(2018)}]{inde2018impact}
\bibinfo{author}{\bibfnamefont{Z.}~\bibnamefont{Inde}} \bibnamefont{and}
  \bibinfo{author}{\bibfnamefont{S.~J.} \bibnamefont{Dixon}},
  \bibinfo{journal}{Critical reviews in biochemistry and molecular biology}
  \textbf{\bibinfo{volume}{53}}, \bibinfo{pages}{99} (\bibinfo{year}{2018}).

\bibitem[{\citenamefont{Chang et~al.}(2008)\citenamefont{Chang, Hemberg,
  Barahona, Ingber, and Huang}}]{chang2008transcriptome}
\bibinfo{author}{\bibfnamefont{H.~H.} \bibnamefont{Chang}},
  \bibinfo{author}{\bibfnamefont{M.}~\bibnamefont{Hemberg}},
  \bibinfo{author}{\bibfnamefont{M.}~\bibnamefont{Barahona}},
  \bibinfo{author}{\bibfnamefont{D.~E.} \bibnamefont{Ingber}},
  \bibnamefont{and} \bibinfo{author}{\bibfnamefont{S.}~\bibnamefont{Huang}},
  \bibinfo{journal}{Nature} \textbf{\bibinfo{volume}{453}},
  \bibinfo{pages}{544} (\bibinfo{year}{2008}).

\bibitem[{\citenamefont{Huang et~al.}(2005)\citenamefont{Huang, Eichler,
  Bar-Yam, and Ingber}}]{huang2005cell}
\bibinfo{author}{\bibfnamefont{S.}~\bibnamefont{Huang}},
  \bibinfo{author}{\bibfnamefont{G.}~\bibnamefont{Eichler}},
  \bibinfo{author}{\bibfnamefont{Y.}~\bibnamefont{Bar-Yam}}, \bibnamefont{and}
  \bibinfo{author}{\bibfnamefont{D.~E.} \bibnamefont{Ingber}},
  \bibinfo{journal}{Physical review letters} \textbf{\bibinfo{volume}{94}},
  \bibinfo{pages}{128701} (\bibinfo{year}{2005}).

\bibitem[{\citenamefont{Kaern et~al.}(2005)\citenamefont{Kaern, Elston, Blake,
  and Collins}}]{kaern2005stochasticity}
\bibinfo{author}{\bibfnamefont{M.}~\bibnamefont{Kaern}},
  \bibinfo{author}{\bibfnamefont{T.~C.} \bibnamefont{Elston}},
  \bibinfo{author}{\bibfnamefont{W.~J.} \bibnamefont{Blake}}, \bibnamefont{and}
  \bibinfo{author}{\bibfnamefont{J.~J.} \bibnamefont{Collins}},
  \bibinfo{journal}{Nature Reviews Genetics} \textbf{\bibinfo{volume}{6}},
  \bibinfo{pages}{451} (\bibinfo{year}{2005}).

\bibitem[{\citenamefont{Luria and Delbr{\"u}ck}(1943)}]{luria1943mutations}
\bibinfo{author}{\bibfnamefont{S.~E.} \bibnamefont{Luria}} \bibnamefont{and}
  \bibinfo{author}{\bibfnamefont{M.}~\bibnamefont{Delbr{\"u}ck}},
  \bibinfo{journal}{Genetics} \textbf{\bibinfo{volume}{28}},
  \bibinfo{pages}{491} (\bibinfo{year}{1943}).

\bibitem[{\citenamefont{Kessler et~al.}(2014)\citenamefont{Kessler, Austin, and
  Levine}}]{kessler2014resistance}
\bibinfo{author}{\bibfnamefont{D.~A.} \bibnamefont{Kessler}},
  \bibinfo{author}{\bibfnamefont{R.~H.} \bibnamefont{Austin}},
  \bibnamefont{and} \bibinfo{author}{\bibfnamefont{H.}~\bibnamefont{Levine}},
  \bibinfo{journal}{Cancer research} \textbf{\bibinfo{volume}{74}},
  \bibinfo{pages}{4663} (\bibinfo{year}{2014}).

\bibitem[{\citenamefont{Komarova}(2006)}]{komarova2006stochastic}
\bibinfo{author}{\bibfnamefont{N.}~\bibnamefont{Komarova}},
  \bibinfo{journal}{Journal of theoretical biology}
  \textbf{\bibinfo{volume}{239}}, \bibinfo{pages}{351} (\bibinfo{year}{2006}).

\bibitem[{\citenamefont{Kalluri and Weinberg}(2009)}]{kalluri2009basics}
\bibinfo{author}{\bibfnamefont{R.}~\bibnamefont{Kalluri}} \bibnamefont{and}
  \bibinfo{author}{\bibfnamefont{R.~A.} \bibnamefont{Weinberg}},
  \bibinfo{journal}{The Journal of clinical investigation}
  \textbf{\bibinfo{volume}{119}}, \bibinfo{pages}{1420} (\bibinfo{year}{2009}).

\bibitem[{\citenamefont{Lamouille et~al.}(2014)\citenamefont{Lamouille, Xu, and
  Derynck}}]{lamouille2014molecular}
\bibinfo{author}{\bibfnamefont{S.}~\bibnamefont{Lamouille}},
  \bibinfo{author}{\bibfnamefont{J.}~\bibnamefont{Xu}}, \bibnamefont{and}
  \bibinfo{author}{\bibfnamefont{R.}~\bibnamefont{Derynck}},
  \bibinfo{journal}{Nature reviews Molecular cell biology}
  \textbf{\bibinfo{volume}{15}}, \bibinfo{pages}{178} (\bibinfo{year}{2014}).

\bibitem[{\citenamefont{Heerboth et~al.}(2015)\citenamefont{Heerboth, Housman,
  Leary, Longacre, Byler, Lapinska, Willbanks, and Sarkar}}]{heerboth2015emt}
\bibinfo{author}{\bibfnamefont{S.}~\bibnamefont{Heerboth}},
  \bibinfo{author}{\bibfnamefont{G.}~\bibnamefont{Housman}},
  \bibinfo{author}{\bibfnamefont{M.}~\bibnamefont{Leary}},
  \bibinfo{author}{\bibfnamefont{M.}~\bibnamefont{Longacre}},
  \bibinfo{author}{\bibfnamefont{S.}~\bibnamefont{Byler}},
  \bibinfo{author}{\bibfnamefont{K.}~\bibnamefont{Lapinska}},
  \bibinfo{author}{\bibfnamefont{A.}~\bibnamefont{Willbanks}},
  \bibnamefont{and} \bibinfo{author}{\bibfnamefont{S.}~\bibnamefont{Sarkar}},
  \bibinfo{journal}{Clinical and translational medicine}
  \textbf{\bibinfo{volume}{4}}, \bibinfo{pages}{6} (\bibinfo{year}{2015}).

\bibitem[{\citenamefont{Yang and Weinberg}(2008)}]{yang2008epithelial}
\bibinfo{author}{\bibfnamefont{J.}~\bibnamefont{Yang}} \bibnamefont{and}
  \bibinfo{author}{\bibfnamefont{R.~A.} \bibnamefont{Weinberg}},
  \bibinfo{journal}{Developmental cell} \textbf{\bibinfo{volume}{14}},
  \bibinfo{pages}{818} (\bibinfo{year}{2008}).

\bibitem[{\citenamefont{Zhang and Weinberg}(2018)}]{zhang2018epithelial}
\bibinfo{author}{\bibfnamefont{Y.}~\bibnamefont{Zhang}} \bibnamefont{and}
  \bibinfo{author}{\bibfnamefont{R.~A.} \bibnamefont{Weinberg}},
  \bibinfo{journal}{Frontiers of medicine} pp. \bibinfo{pages}{1--13}
  (\bibinfo{year}{2018}).

\bibitem[{\citenamefont{Thiery et~al.}(2009)\citenamefont{Thiery, Acloque,
  Huang, and Nieto}}]{thiery2009epithelial}
\bibinfo{author}{\bibfnamefont{J.~P.} \bibnamefont{Thiery}},
  \bibinfo{author}{\bibfnamefont{H.}~\bibnamefont{Acloque}},
  \bibinfo{author}{\bibfnamefont{R.~Y.} \bibnamefont{Huang}}, \bibnamefont{and}
  \bibinfo{author}{\bibfnamefont{M.~A.} \bibnamefont{Nieto}},
  \bibinfo{journal}{cell} \textbf{\bibinfo{volume}{139}}, \bibinfo{pages}{871}
  (\bibinfo{year}{2009}).

\bibitem[{\citenamefont{Singh and Settleman}(2010)}]{singh2010emt}
\bibinfo{author}{\bibfnamefont{A.}~\bibnamefont{Singh}} \bibnamefont{and}
  \bibinfo{author}{\bibfnamefont{J.}~\bibnamefont{Settleman}},
  \bibinfo{journal}{Oncogene} \textbf{\bibinfo{volume}{29}},
  \bibinfo{pages}{4741} (\bibinfo{year}{2010}).

\bibitem[{\citenamefont{Lu et~al.}(2013)\citenamefont{Lu, Jolly, Levine,
  Onuchic, and Ben-Jacob}}]{lu2013microrna}
\bibinfo{author}{\bibfnamefont{M.}~\bibnamefont{Lu}},
  \bibinfo{author}{\bibfnamefont{M.~K.} \bibnamefont{Jolly}},
  \bibinfo{author}{\bibfnamefont{H.}~\bibnamefont{Levine}},
  \bibinfo{author}{\bibfnamefont{J.~N.} \bibnamefont{Onuchic}},
  \bibnamefont{and}
  \bibinfo{author}{\bibfnamefont{E.}~\bibnamefont{Ben-Jacob}},
  \bibinfo{journal}{Proceedings of the National Academy of Sciences} p.
  \bibinfo{pages}{201318192} (\bibinfo{year}{2013}).

\bibitem[{\citenamefont{Jolly et~al.}(2015)\citenamefont{Jolly, Boareto, Huang,
  Jia, Lu, Ben-Jacob, Onuchic, and Levine}}]{jolly2015implications}
\bibinfo{author}{\bibfnamefont{M.~K.} \bibnamefont{Jolly}},
  \bibinfo{author}{\bibfnamefont{M.}~\bibnamefont{Boareto}},
  \bibinfo{author}{\bibfnamefont{B.}~\bibnamefont{Huang}},
  \bibinfo{author}{\bibfnamefont{D.}~\bibnamefont{Jia}},
  \bibinfo{author}{\bibfnamefont{M.}~\bibnamefont{Lu}},
  \bibinfo{author}{\bibfnamefont{E.}~\bibnamefont{Ben-Jacob}},
  \bibinfo{author}{\bibfnamefont{J.~N.} \bibnamefont{Onuchic}},
  \bibnamefont{and} \bibinfo{author}{\bibfnamefont{H.}~\bibnamefont{Levine}},
  \bibinfo{journal}{Frontiers in oncology} \textbf{\bibinfo{volume}{5}},
  \bibinfo{pages}{155} (\bibinfo{year}{2015}).

\bibitem[{\citenamefont{Jolly et~al.}(2016)\citenamefont{Jolly, Tripathi, Jia,
  Mooney, Celiktas, Hanash, Mani, Pienta, Ben-Jacob, and
  Levine}}]{jolly2016stability}
\bibinfo{author}{\bibfnamefont{M.~K.} \bibnamefont{Jolly}},
  \bibinfo{author}{\bibfnamefont{S.~C.} \bibnamefont{Tripathi}},
  \bibinfo{author}{\bibfnamefont{D.}~\bibnamefont{Jia}},
  \bibinfo{author}{\bibfnamefont{S.~M.} \bibnamefont{Mooney}},
  \bibinfo{author}{\bibfnamefont{M.}~\bibnamefont{Celiktas}},
  \bibinfo{author}{\bibfnamefont{S.~M.} \bibnamefont{Hanash}},
  \bibinfo{author}{\bibfnamefont{S.~A.} \bibnamefont{Mani}},
  \bibinfo{author}{\bibfnamefont{K.~J.} \bibnamefont{Pienta}},
  \bibinfo{author}{\bibfnamefont{E.}~\bibnamefont{Ben-Jacob}},
  \bibnamefont{and} \bibinfo{author}{\bibfnamefont{H.}~\bibnamefont{Levine}},
  \bibinfo{journal}{Oncotarget} \textbf{\bibinfo{volume}{7}},
  \bibinfo{pages}{27067} (\bibinfo{year}{2016}).

\bibitem[{\citenamefont{Hong et~al.}(2015)\citenamefont{Hong, Watanabe, Ta,
  Villarreal-Ponce, Nie, and Dai}}]{hong2015ovol2}
\bibinfo{author}{\bibfnamefont{T.}~\bibnamefont{Hong}},
  \bibinfo{author}{\bibfnamefont{K.}~\bibnamefont{Watanabe}},
  \bibinfo{author}{\bibfnamefont{C.~H.} \bibnamefont{Ta}},
  \bibinfo{author}{\bibfnamefont{A.}~\bibnamefont{Villarreal-Ponce}},
  \bibinfo{author}{\bibfnamefont{Q.}~\bibnamefont{Nie}}, \bibnamefont{and}
  \bibinfo{author}{\bibfnamefont{X.}~\bibnamefont{Dai}}, \bibinfo{journal}{PLoS
  computational biology} \textbf{\bibinfo{volume}{11}},
  \bibinfo{pages}{e1004569} (\bibinfo{year}{2015}).

\bibitem[{\citenamefont{Li and Balazsi}(2018)}]{li2018landscape}
\bibinfo{author}{\bibfnamefont{C.}~\bibnamefont{Li}} \bibnamefont{and}
  \bibinfo{author}{\bibfnamefont{G.}~\bibnamefont{Balazsi}},
  \bibinfo{journal}{NPJ systems biology and applications}
  \textbf{\bibinfo{volume}{4}}, \bibinfo{pages}{34} (\bibinfo{year}{2018}).

\bibitem[{\citenamefont{Collisson et~al.}(2011)\citenamefont{Collisson,
  Sadanandam, Olson, Gibb, Truitt, Gu, Cooc, Weinkle, Kim, Jakkula
  et~al.}}]{collisson2011subtypes}
\bibinfo{author}{\bibfnamefont{E.~A.} \bibnamefont{Collisson}},
  \bibinfo{author}{\bibfnamefont{A.}~\bibnamefont{Sadanandam}},
  \bibinfo{author}{\bibfnamefont{P.}~\bibnamefont{Olson}},
  \bibinfo{author}{\bibfnamefont{W.~J.} \bibnamefont{Gibb}},
  \bibinfo{author}{\bibfnamefont{M.}~\bibnamefont{Truitt}},
  \bibinfo{author}{\bibfnamefont{S.}~\bibnamefont{Gu}},
  \bibinfo{author}{\bibfnamefont{J.}~\bibnamefont{Cooc}},
  \bibinfo{author}{\bibfnamefont{J.}~\bibnamefont{Weinkle}},
  \bibinfo{author}{\bibfnamefont{G.~E.} \bibnamefont{Kim}},
  \bibinfo{author}{\bibfnamefont{L.}~\bibnamefont{Jakkula}},
  \bibnamefont{et~al.}, \bibinfo{journal}{Nature medicine}
  \textbf{\bibinfo{volume}{17}}, \bibinfo{pages}{500} (\bibinfo{year}{2011}).

\bibitem[{\citenamefont{Antal and Krapivsky}(2011)}]{antal2011exact}
\bibinfo{author}{\bibfnamefont{T.}~\bibnamefont{Antal}} \bibnamefont{and}
  \bibinfo{author}{\bibfnamefont{P.}~\bibnamefont{Krapivsky}},
  \bibinfo{journal}{Journal of Statistical Mechanics: Theory and Experiment}
  \textbf{\bibinfo{volume}{2011}}, \bibinfo{pages}{P08018}
  (\bibinfo{year}{2011}).

\bibitem[{\citenamefont{Gillespie}(1977)}]{gillespie1977exact}
\bibinfo{author}{\bibfnamefont{D.~T.} \bibnamefont{Gillespie}},
  \bibinfo{journal}{The journal of physical chemistry}
  \textbf{\bibinfo{volume}{81}}, \bibinfo{pages}{2340} (\bibinfo{year}{1977}).

\bibitem[{\citenamefont{Toneff et~al.}(2016)\citenamefont{Toneff, Sreekumar,
  Tinnirello, Den~Hollander, Habib, Li, Ellis, Xin, Mani, and
  Rosen}}]{toneff2016z}
\bibinfo{author}{\bibfnamefont{M.}~\bibnamefont{Toneff}},
  \bibinfo{author}{\bibfnamefont{A.}~\bibnamefont{Sreekumar}},
  \bibinfo{author}{\bibfnamefont{A.}~\bibnamefont{Tinnirello}},
  \bibinfo{author}{\bibfnamefont{P.}~\bibnamefont{Den~Hollander}},
  \bibinfo{author}{\bibfnamefont{S.}~\bibnamefont{Habib}},
  \bibinfo{author}{\bibfnamefont{S.}~\bibnamefont{Li}},
  \bibinfo{author}{\bibfnamefont{M.}~\bibnamefont{Ellis}},
  \bibinfo{author}{\bibfnamefont{L.}~\bibnamefont{Xin}},
  \bibinfo{author}{\bibfnamefont{S.}~\bibnamefont{Mani}}, \bibnamefont{and}
  \bibinfo{author}{\bibfnamefont{J.}~\bibnamefont{Rosen}},
  \bibinfo{journal}{BMC biology} \textbf{\bibinfo{volume}{14}},
  \bibinfo{pages}{47} (\bibinfo{year}{2016}).

\bibitem[{\citenamefont{Cramer et~al.}(2017)\citenamefont{Cramer, Jones,
  El-Hamidi, and Celli}}]{cramer2017ecm}
\bibinfo{author}{\bibfnamefont{G.~M.} \bibnamefont{Cramer}},
  \bibinfo{author}{\bibfnamefont{D.~P.} \bibnamefont{Jones}},
  \bibinfo{author}{\bibfnamefont{H.}~\bibnamefont{El-Hamidi}},
  \bibnamefont{and} \bibinfo{author}{\bibfnamefont{J.~P.} \bibnamefont{Celli}},
  \bibinfo{journal}{Molecular Cancer Research} \textbf{\bibinfo{volume}{15}},
  \bibinfo{pages}{15} (\bibinfo{year}{2017}).

\end{thebibliography}
\newpage

{\hspace*{6cm}{{\bf{\Large Supplementary Material}}}

\maketitle
\section*{Supplementary Material A: Analytical results for surviving population upon exposure to drugs}
\setcounter{equation}{0}
\renewcommand{\theequation}{A\arabic{equation}}
\subsection*{Derivation of single cell generating function}
Consider a single cell that can exist either as an epithelial($E$) or mesenchymal($M$) cell. The rates of switching between these two phenotypes ($E$
and $M$) are given by  $k_{EM}$ and $k_{ME}$ and the rates of cell death are given by
$\mu_E$ and $\mu_M$, respectively. Here we assume no cell divisions (i.e no new production of cells) for  either $E$ or $M$ due to high levels of 
external drugs. For such a system, we consider first the temporal evolution of a single cell, given the initial probability $p_0$ of the cell being $M$-type.  The corresponding probability generating function is $g(z_1,z_2,t | p_0)$
$=\sum_{\eta_E}\sum_{\eta_M}z_1^{\eta_E}z_2^{\eta_M}P(\eta_E,\eta_M,t | p_0)$, where $\eta_E$ and $\eta_M$ can have values 
0 or 1, and $P(\eta_E,\eta_M,t | p_0)$ is the probability of having $\eta_E$ and $\eta_M$ number of cells
at time $t$, given the initial proability $p_0$ of the cell being $M$-type. Clearly $P(1,1,t| p_0) =0$ since we are starting with a single cell and no new cells are created. Correspondingly at any time $t$, we have only three
possibilities, either $\eta_E=1$ and $\eta_M=0$ or $\eta_E=0$ and $\eta_M=1$ or $\eta_E=\eta_M=0$. Thus we obtain
\begin{equation}{\label{eA1}}
g(z_1,z_2,t|p_0)=P(0,0,t|p_0)+z_1P(1,0,t|p_0)+z_2P(0,1,t|p_0).
\end{equation}
Denote $P(1,0,t|p_0)=P_E(t)$, $P(0,1,t|p_0)=P_M(t)$ and $P(0,0,t|p_0)=P_0(t)$,  and then using the normalization condition, $P_E(t)+P_M(t)+P_0(t)=1$, 
we obtain the single particle generating function as derived in the main text, Eq. (2).\\

To find an explicit expression for the single cell generating function $g(z_1,z_2,t|p_0)$, we need to find
expressions for the probabilities $P_E(t)$ and $P_M(t)$. For this, we begin from their evolution equations:
\begin{eqnarray}{\label{eA2}}
\frac{dP_E(t)}{dt}&=&k_{ME}P_M(t)-(\mu_E+k_{EM})P_M(t),\nonumber\\
\frac{dP_M(t)}{dt}&=&k_{EM}P_E(t)-(\mu_M+k_{ME})P_M(t).\nonumber\\
\end{eqnarray}
These equations can be solved to give the following expressions for the temporal evolution of the probabilities:
 \begin{eqnarray}{\label{eA5}}
 P_E&=& \left[\frac{(1-p_0)(\gamma_0+\alpha_0-2\mu_M)-2k_{ME}}{2\alpha}\right]\exp\left(-\frac{t}{2}(\gamma_0+\alpha_0)\right)
           -\left[\frac{(1-p_0)(\gamma_0-\alpha_0-2\mu_M)-2k_{ME}}{2\alpha}\right]\exp\left(-\frac{t}{2}(\gamma_0-\alpha_0)\right),\nonumber\\
 P_M&=&\left[\frac{p_0(\gamma_0+\alpha_0-2\mu_E)-2k_{EM}}{2\alpha}\right]\exp\left(-\frac{t}{2}(\gamma_0+\alpha_0)\right)
           -\left[\frac{p_0(\gamma_0-\alpha_0-2\mu_E)-2k_{EM}}{2\alpha}\right]\exp\left(-\frac{t}{2}(\gamma_0-\alpha_0)\right),
\end{eqnarray}

where $\alpha_0$ and $\gamma_0$ denote combinations of the model parameters, as given in Eq.(5) in the main text and $p_0=P_M(t=0)$ as discussed above. Using the
expressions for $P_E$ and $P_M$ from Eq. (\ref{eA5}) in Eq. (2), we obtain the single particle probability generating function.\\


\subsection*{Derivation of mean and Fano factor for surviving population}
Here we provide the details for the derivation of expressions for temporal evolution of moments associated with total surviving population. For this, we 
start from the expression for the generating function, 
\begin{equation}{\label{eA6}}
G(z_1,z_2,t)=\sum_{E=0}^{\infty}\sum_{M=0}^{\infty}z_1^Ez_2^MP(E,M,t)=\int_{p_0=0}^{p_0=1}dp_0\rho(p_0)
\left[g(z_1,z_2,t)\right]^{N_0},
\end{equation}
and using Eqs.(\ref{eA6}), (2), (\ref{eA5}) and denoting $\langle p_0\rangle=\int p_0\rho(p_0)dp_0$ as the mean value of $p_0$, arrive at the following expressions for the mean numbers of epithelial and mesenchymal cells,
\begin{eqnarray}{\label{eA8}}
\langle E\rangle&=&\left. \frac{dG}{dz_1}\right |_{1,1}=N_0(S_0+S_1\langle p_0\rangle),\nonumber \\
\langle M \rangle &=& \langle M \rangle=\left. \frac{dG}{dz_2} \right |_{1,1}= N_0(Q_0+Q_1 \langle p_0\rangle),\nonumber\\
\end{eqnarray}
where
\begin{eqnarray}{\label{eA9}}
Q_0&=&\left(\frac{-2k_{EM}}{2\alpha_0}\right)\left(\exp\left(-\frac{t}{2}(\gamma_0+\alpha_0)\right)-\exp\left(-\frac{t}{2}(\gamma_0-\alpha_0)\right)\right),\nonumber\\
Q_1&=&\left(\frac{\gamma_0+\alpha_0-2\mu_E}{2\alpha_0}\right)\exp\left(-\frac{t}{2}(\gamma_0+\alpha_0)\right)-\left(\frac{\gamma_0-\alpha_0-2\mu_E}{2\alpha_0}\right)\exp\left(-\frac{t}{2}(\gamma_0-\alpha_0)\right),\nonumber \\
S_0&=&\left(\frac{\gamma_0+\alpha_0-2(k_{ME}+\mu_M)}{2\alpha_0}\right)\exp\left(-\frac{t}{2}(\gamma_0+\alpha_0)\right)-\left(\frac{\gamma_0-\alpha_0-2(k_{ME}+\mu_M)}{2\alpha_0}\right)
\exp\left(-\frac{t}{2}(\gamma_0-\alpha_0)\right), \nonumber\\
S_1&=&\left(\frac{\gamma_0-\alpha_0-2\mu_M}{2\alpha_0}\right)\exp\left(-\frac{t}{2}(\gamma_0-\alpha_0)\right)-\left(\frac{\gamma_0+\alpha_0-2\mu_M}{2\alpha_0}\right)\exp\left(-\frac{t}{2}(\gamma_0+\alpha_0)\right),
\end{eqnarray}
 and $\alpha_0$ and $\gamma_0$ are given by Eq. (5) in the main text. Using Eq.(\ref{eA8}), we see that the 
mean number of total surviving cells ($E+M$) is given by
\begin{equation}{\label{eA10}}
\langle N\rangle=N_0\left[Q_0+S_0+(Q_1+S_1)\langle p_0 \rangle\right].
\end{equation}
We turn next to the second moment and, using the same generating function, the variances in $E$ and $M$ cells are given by
\begin{eqnarray}{\label{eA11}}
\sigma^2_E=\left. \frac{d^2G}{dz_1^2}\right |_{1,1}-\left(\left. \frac{dG}{dz_1}\right |_{1,1}\right)^2+\left. \frac{dG}{dz_1}\right |_{1,1},\nonumber\\
\sigma^2_M=\left. \frac{d^2G}{dz_2^2}\right |_{1,1}-\left(\left. \frac{dG}{dz_2}\right |_{1,1}\right)^2+\left. \frac{dG}{dz_2}\right |_{1,1},
\end{eqnarray}
leading to the following expression for the Fano factor:
\begin{eqnarray}{\label{eA12}}
 F_E&=&\frac{\sigma^2_E}{\langle E \rangle}=1-\frac{\langle E\rangle}{N_0}+\frac{N_0(N_0-1)}{\langle E\rangle}S_1^2\sigma^2_{p_0},\nonumber \\
 F_M&=&\frac{\sigma^2_M}{\langle M \rangle}=1-\frac{\langle M\rangle}{N_0}+\frac{N_0(N_0-1)}{\langle M\rangle}Q_1^2\sigma^2_{p_0}.
\end{eqnarray}
Substituting the expressions for  $S_1$ and $Q_1$ and simplifying these Fano factors can be reexpressed
as
\begin{eqnarray}{\label{eA13}}
 F_E&=&1-\frac{\langle E \rangle}{N_0}+\frac{N_0(N_0-1)}{\langle E \rangle}\left[\left(\frac{\gamma_0-\alpha_0-2\mu_M}{2\alpha_0}\right)\exp\left(-\frac{t}{2}(\gamma_0-\alpha_0)\right)
  -\left(\frac{\gamma_0+\alpha_0-2\mu_M}{2\alpha_0}\right)\exp\left(-\frac{t}{2}(\gamma_0+\alpha_0)\right)\right]^2\sigma^2_{p_0},\nonumber\\
 F_M&=&1-\frac{\langle M \rangle}{N_0}+\frac{N_0(N_0-1)}{\langle M \rangle}\left[\left(\frac{\gamma_0+\alpha_0-2\mu_M}{2\alpha_0}\right)\exp\left(-\frac{t}{2}(\gamma_0+\alpha_0)\right)
 -\left(\frac{\gamma_0-\alpha_0-2\mu_M}{2\alpha_0}\right)\exp\left(-\frac{t}{2}(\gamma_0-\alpha_0)\right)\right]^2\sigma^2_{p_0}.\nonumber\\
\end{eqnarray}
Note that both the Fano factors, $F_{E}$ and $F_{M}$, are always less than one if $\sigma^2_{p_0}=0$. That is, a Fano factor greater than 1 is an indication of the presence of initial variability in the fraction of $M$ cells in the population.\\

To derive an expression for the variance of the total population $N=E+M$, we use the relation 
$\sigma^2_N=\sigma^2_E+\sigma^2_M+2C_{EM}$, where $C_{EM}=\langle EM \rangle-\langle E \rangle \langle M \rangle$ is the correlation between $E$ and $M$, and $\langle EM \rangle$, in terms 
of the generating function, is
\begin{equation}{\label{eA14}}
\langle EM \rangle=\left. \frac{d}{dz_1}\left(\frac{dG}{dz_2}\right)\right|_{1,1}.
\end{equation}
The expression for the Fano factor for the total population $N$ can be written as
\begin{equation}{\label{eA15}}
F=1-\frac{\langle N \rangle}{N_0}+\frac{N_0(N_0-1)}{\langle N \rangle}(S_1+Q_1)^2\sigma^2_{p_0},
\end{equation}
which, on using Eq. (\ref{eA9}),  can be rewritten as Eq. (7) in the main text.

\subsection*{Expressions for $p_0$ and $\sigma^2_{p_0}$}
 As discussed in the main text, we propose a three-step procedure. 
First, we set $p_0=0$, which, using Eqs. (\ref{eA10}) and (\ref{eA15}), gives
\begin{eqnarray}{\label{eA16}}
 \langle N \rangle_0&=&N_0(Q_0+S_0),\nonumber \\
 F_0&=&1-(Q_0+S_0),
\end{eqnarray}
and next, setting  $p_0=1$,
\begin{eqnarray}{\label{eA17}}
 \langle N \rangle_1&=&N_0(Q_0+S_0+Q_1+S_1),\nonumber \\
 F_0&=&1-(Q_0+S_0+Q_1+S_1).
\end{eqnarray}
Finally, using the expressions for mean number of surviving cells, Eq.(\ref{eA10}), and corresponding expression for the 
Fano factor, Eq. (\ref{eA15}), for arbitrary $p_0$, we get explicit expressions for the probability $p_0$
and variance in initial M-type cells $\sigma^2_{p_0}$. The resulting expressions are in terms of mean values  $\langle N \rangle_0$ 
(for $p_0=0$) and  $\langle N \rangle_1$ (for $p_0=1$) and the mean and Fano factor
for a given arbitrary $p_0$, as shown in the main text.


\section*{Supplementary Material B: Analytical results in the growth phase}
\setcounter{equation}{0}
\renewcommand{\theequation}{B\arabic{equation}}
In this section, we provide details of the derivation of the moments for the surviving cell 
populations in the growth phase.\\

\subsection*{ First moments} Multiplying Eq. (1) by E or M and summing over all possible values of E and M, we obtain the evolution 
equations for $\langle E \rangle=\sum E P(E,M,t)$ and $\langle M \rangle=\sum M P(E,M,t)$ :  
\begin{eqnarray}{\label{eB1}}
 \frac{\partial \langle E \rangle}{\partial t}&=&\left(k_E-\mu_E-k_{EM}\right)\langle E \rangle+k_{ME}\langle M\rangle,\nonumber\\
\frac{\partial \langle M \rangle}{\partial t}&=& \left(k_M-\mu_M-k_{ME}\right)\langle M \rangle+k_{EM}\langle E \rangle,
\end{eqnarray}
These equations can be solved to get:
\begin{eqnarray}{\label{eB4}}
 \langle E \rangle&=&\frac{1}{2\alpha}\left[(E_0\alpha-\beta_E)\exp\left(-\frac{t}{2}(\gamma+\alpha)\right)+(E_0\alpha+\beta_E)\exp\left(-\frac{t}{2}(\gamma-\alpha)\right)\right],\nonumber\\
 \langle M \rangle &=&\frac{1}{2\alpha}\left[(M_0\alpha-\beta_M)\exp\left(-\frac{t}{2}(\gamma+\alpha)\right)+(M_0\alpha+\beta_M)\exp\left(-\frac{t}{2}(\gamma-\alpha)\right)\right],
\end{eqnarray}

where $E_0$ and $M_0$ are  initial values for the number of $E$-type and $M$-type cells respectively, and 
\begin{eqnarray}{\label{eB5}}
 \gamma&=&k_{EM}+k_{ME}-k_E^f-k_M^f,\nonumber\\
 \alpha&=&\sqrt{\gamma^2+4\left(k_{ME}(k_E^f-k_M^f)+(\gamma+k_M^f)k_M^f\right)},\nonumber\\
 \beta_E&=&2M_0k_{ME}+E_0\left(\gamma-2(k_{EM}-k_E^f)\right),\nonumber\\
 \beta_M&=&2E_0k_{EM}+M_0\left(\gamma-2(k_{ME}-k_M^f)\right)
\end{eqnarray}
 with 
\begin{equation}{\label{eB6}}
k_E^f=k_E-\mu_E, ~~~k_M^f=k_M-\mu_M,
\end{equation}
representing the effective birth rates for $E$-type and $M$-type cells, respectively. 
For the total population, $N=E+M$, using Eq. (\ref{eB4}), the mean $\langle N\rangle=$ $\langle E\rangle+\langle M \rangle$ can then be written as:
\begin{eqnarray}{\label{eBN}}
 \langle N \rangle&=&\left(\frac{E_0(\alpha-\gamma-2k_E^f)+M_0(\alpha-\gamma-2k_M^f)}{2\alpha}\right)\exp\left(-\frac{t}{2}(\gamma+\alpha)\right)\nonumber\\
 &+&\left(\frac{E_0(\alpha+\gamma+2k_E^f)+M_0(\alpha+\gamma+2k_M^f)}{2\alpha}\right)\exp\left(-\frac{t}{2}(\gamma-\alpha)\right).
\end{eqnarray}


\subsection* {Second moments} Using Eq. (1), we can write the evolution equation for 
$$\langle E^2 \rangle=\sum E^2 P(E,M,t), ~~\langle M^2 \rangle=\sum M^2 P(E,M,t)~~ \text{and} ~~\langle EM \rangle=\sum EM P(E,M,t)$$ as:
\begin{eqnarray}{\label{eB3-1}}
 \frac{\partial \langle E^2\rangle}{\partial t}&=&(k_E+\mu_E+k_{EM}) \langle E \rangle+k_{ME}\langle M \rangle+2(k_E-\mu_E-k_{EM})\langle E^2\rangle+2k_{ME}\langle ME \rangle,\nonumber\\
\frac{\partial \langle M^2\rangle}{\partial t}&=&(k_M+\mu_M+k_{ME})\langle M \rangle+k_{EM}\langle E \rangle+2(k_M-\mu_M-k_{ME})\langle M^2\rangle+2k_{EM}\langle ME \rangle,\nonumber\\
\frac{\partial \langle ME \rangle}{\partial t}&=&-k_{ME}\langle M\rangle-k_{EM}\langle E \rangle+(k_E+k_M-\mu_E-\mu_M-k_{ME}-k_{EM})\langle ME \rangle +k_{ME}\langle M^2\rangle
+k_{EM}\langle E^2\rangle.
\end{eqnarray}
Eliminating variables in Eq. (\ref{eB3-1}), we obtain a single ODE for $\langle E^2\rangle$,

\begin{equation}{\label{eB3-2}}
\frac{d^3\langle E^2 \rangle}{dt^3}-c_2\frac{d^2\langle E^2 \rangle}{dt^2}+c_1\frac{d\langle E^2 \rangle}{dt}
+c_0\langle E^2 \rangle=\mathcal{K}(t),
\end{equation}
where $c_0$, $c_1$ and $c_2$ are functions of the model parameters
\begin{eqnarray}{\label{eB3-3}}
 c_2&=&3(k_E+k_M-k_{EM}-k_{ME}-\mu_E-\mu_M),\nonumber\\
 c_1&=&2\left[k_E^2+k_{EM}^2+k_M^2-2k_Mk_{ME}+k_{ME^2}-4k_M\mu_E+4k_{ME}\mu_e+\mu_E^2-2k_m\mu_M+2k_{ME}\mu_M+4\mu_e\mu_M+\mu_M^2\right.\nonumber\\ &+&\left. 2k_{EM}(-2k_M+k_{ME}+\mu_E+2\mu_M)
-2k_E(k_{EM}-2k_M+2k_{ME}+\mu_E+2\mu_M)\right],\nonumber\\
c_0&=&4(k_E+k_M-k_{EM}-k_{ME}-\mu_E-\mu_M)\left[k_{EM}k_M+k_M\mu_E-k_{ME}\mu_E-(k_{EM}+\mu_E)\mu_M+k_E(k_{ME}+\mu_M-k_M)\right],\nonumber\\
\end{eqnarray}
and $\mathcal{K}(t)$ is a known function in $t$:
\begin{eqnarray}{\label{eB3-4}}
\mathcal{K}(t)&=&\frac{1}{2k_{ME}^2}\left[(k_M+\mu_M+k_{ME}-k_{EM})\langle M \rangle+k_{EM}\left( 1-\frac{k_E+\mu_E+k_{EM}}{k_{ME}}\right)\langle E\rangle 
+2(k_M-\mu_M-k_ME)g(t)-\frac{dg(t)}{dt}\right],\nonumber\\
\end{eqnarray}
with
\begin{eqnarray}{\label{eB3-5}}
 g(t)&=&-\frac{1}{2k_{ME}^2}\left[(k_E+\mu_E+k_{EM})\frac{d\langle E \rangle}{dt}+k_{ME}\frac{d\langle M \rangle}{dt}\right]+\langle M \rangle +\frac{k_{EM}}{k_{ME}}\langle E \rangle
+\frac{k_E+k_M-\mu_E-\mu_M-k_{ME}-k_{EM}}{2k_{ME}^2}\nonumber\\&&\left[(k_E+\mu_E+k_{EM})\langle E \rangle+k_{ME}\langle M\rangle\right].
\end{eqnarray}
The solution for $\langle E^2 \rangle$ in Eq. (\ref{eB3-2}) is:
\begin{eqnarray}{\label{eB3-6}}
 \langle E^2 \rangle&=&\sum_{i=1}^3 s_i\exp[\lambda_it]+\sum_{i=1}^3\frac{\exp[\lambda_it]}{3\lambda_i^2-2c_2\lambda_i+c_1}\int{\mathcal{K}(t)\exp(-\lambda_i t)}dt
\end{eqnarray}
where $\lambda_i$, $i=1,2,3$  are the roots of 
\begin{equation}{\label{eB3-7}}
 \lambda^3-c_2\lambda^2+c_1\lambda+c_0=0
\end{equation}
 and $s_1,s_2,s_3$ are constants to be determined from the initial condition. 
Once we have obtained an expression for  $\langle E^2 \rangle$, we can readily derive expressions for  $\langle M^2 \rangle$ and 
 $ \langle ME \rangle$:
\begin{eqnarray}{\label{eB3-11}}
 \langle M^2\rangle&=&\frac{1}{2k_{ME}^2}\left[\frac{d^2\langle E^2\rangle}{dt^2}+(3k_{EM}+k_{ME}-3k_E-k_M+3\mu_E+\mu_M)\frac{d\langle E^2\rangle}{dt}\right. \nonumber\\ &+& \left.
       2\left((k_E-k_{EM}-\mu_E)(k_E-k_{EM}+k_M-k_{ME}-\mu_E-\mu_M)-k_{EM}k_{ME}\right)\langle E^2 \rangle +2k_{ME}^2g(t)\right],\nonumber\\
\langle ME\rangle &=& \frac{1}{2k_{ME}}\left[\frac{d\langle E^2 \rangle}{dt}-\langle M \rangle k_{ME}-\langle E\rangle(k_E+k_{EM}+\mu_E)-2(k_E-k_{EM}-\mu_E)\langle E^2\rangle \right].
\end{eqnarray}
 Taking suitable initial conditions, we can find the constants $s_1,s_2,s_3$ that will help specify the temporal evolution of the second moments in terms of the model parameters. 

\end{document}